\documentclass[aps,prd,superscriptaddress,showkeys,longbibliography,
nofootinbib]{revtex4-2}

\usepackage[english]{babel}
\usepackage{enumitem}

\usepackage{hyperref}
\usepackage{physics}
\usepackage{mathrsfs}
\usepackage{amsmath}
\usepackage{amsfonts}
\usepackage{amssymb}
\usepackage{listings}
\usepackage{mathtools}
\usepackage{graphicx}
\usepackage{amssymb}
\usepackage{bbm}
\usepackage{float}
\usepackage{wrapfig}
\usepackage{array}
\usepackage{cleveref}
\Crefname{figure}{Fig.}{Figs.}
\Crefname{equation}{Eq.}{Eqs.}
\crefname{equation}{}{}

\usepackage{ragged2e}
\usepackage{tcolorbox}

\usepackage{amsmath}
\usepackage{graphicx}

\newcommand{\mn}{{\mu\nu}}
\newcommand{\ab}{{\alpha\beta}}
\newcommand{\m}{\mathfrak{m}}
\newcommand{\intp}{\int_{\textbf{p}}}

\definecolor{applegreen}{rgb}{0.0,0.7,0.0}

\newcommand{\Ip}{\mathcal{I}}
\newcommand{\Dsh}{D_\text{sh}}
\newcommand{\jsh}{J_\text{sh}}
\newcommand{\wRe}{\mathrm{Re}\,\omega}
\newcommand{\Gy}{\mathcal{G}}
\newcommand{\gam}{\xi}

\begin{document}
\title{Relaxation for massive particles: transport and causality}

\author{Matej Bajec}
\email[]{matej.bajec@fmf.uni-lj.si}
\author{Alexander Soloviev}
\email[]{alexander.soloviev@fmf.uni-lj.si}
\affiliation{Faculty of Mathematics and Physics, University of Ljubljana,
Jadranska ulica 19, SI-1000 Ljubljana, Slovenia}

\date{\today}

\begin{abstract}
    Correlators at finite density and temperature encode important information about a physical system, such as transport and causality. For a massive gas of particles in the relaxation time approximation of kinetic theory, we provide complete analytic results for all first order transport coefficients, including thermoelectric coefficients. We demonstrate the first complete picture of the complex structure of the correlators as a function of mass, providing an interpretation in terms of the causal structure of lightcones. We provide a simple criterion to extract the effective lightcone velocity from the cut of the retarded correlator. 
\end{abstract}

\maketitle
\tableofcontents

\section{Introduction}

The emergence of collective phenomena from microscopic dynamics has been a key research frontier, leading to robust characterization of near- and non-equilibrium physics. To that end, an often utilized 
microscopic description is 
kinetic theory \cite{pitaevskii2012physical,groot-book}, which describes the ballistic trajectories of (quasi-)particles under the influence of external fields and scattering against other microscopic degrees of freedom.

Over the years, a crucial theoretical laboratory in kinetic theory has been the relaxation time approximation (RTA) \cite{ANDERSON1974466}. RTA kinetic theory dramatically simplifies the collision kernel by focusing on a single relaxation time for which the system relaxes to equilibrium. As such, it captures the near equilibrium behavior of a gas, while remaining analytically tractable. This is a valid approximation when there is a well-defined lowest eigenvalue of the collision kernel \cite{liboff-book,Grozdanov:2016vgg}. There have been numerous recent results including:  the determination of retarded correlators due to charge and momentum transport \cite{Romatschke:2015gic}, the exact analytic solution in $0+1$ dimensions for transversely homogeneous boost-invariant systems \cite{Florkowski:2013lza,Florkowski:2013lya} (and additionally the massive case \cite{Florkowski:2014sfa}), extensions to energy dependent relaxation time \cite{Kurkela:2017xis,Kurkela:2019kip,Dash:2021ibx,Ambrus:2022koq,Brants:2024wrx, Hu:2024tnn,Singh:2025wov,Mukherjee:2025dqp}, analytic analysis of the Bjorken attractor \cite{Blaizot:2021cdv,Aniceto:2024pyc}, determining the dissipative fluid dynamics from the massive RTA gas 
\cite{Ambrus:2022vif,Jaiswal:2022udf,Ambrus:2023qcl}, serving as a useful benchmark for numerical methods \cite{Ochsenfeld:2023wxz}, the extraction of thermoelectric coefficients from analytic retarded correlators in systems with momentum relaxation \cite{Bajec:2024jez}, analytic determination of the three point function in the Schwinger-Keldysh formalism \cite{Abbasi:2024pwz} and developing a multi-species RTA \cite{Rocha:2025rkl}. Phenomenological applications vary from e.g.~the hadron resonance gas \cite{Dusling:2011fd,Das:2020beh} 
to the Nernst effect in band structures of semiconductors \cite{PhysRevB.103.144404}.

\begin{figure}[h!]
    \centering
    \includegraphics[width=0.8\linewidth]{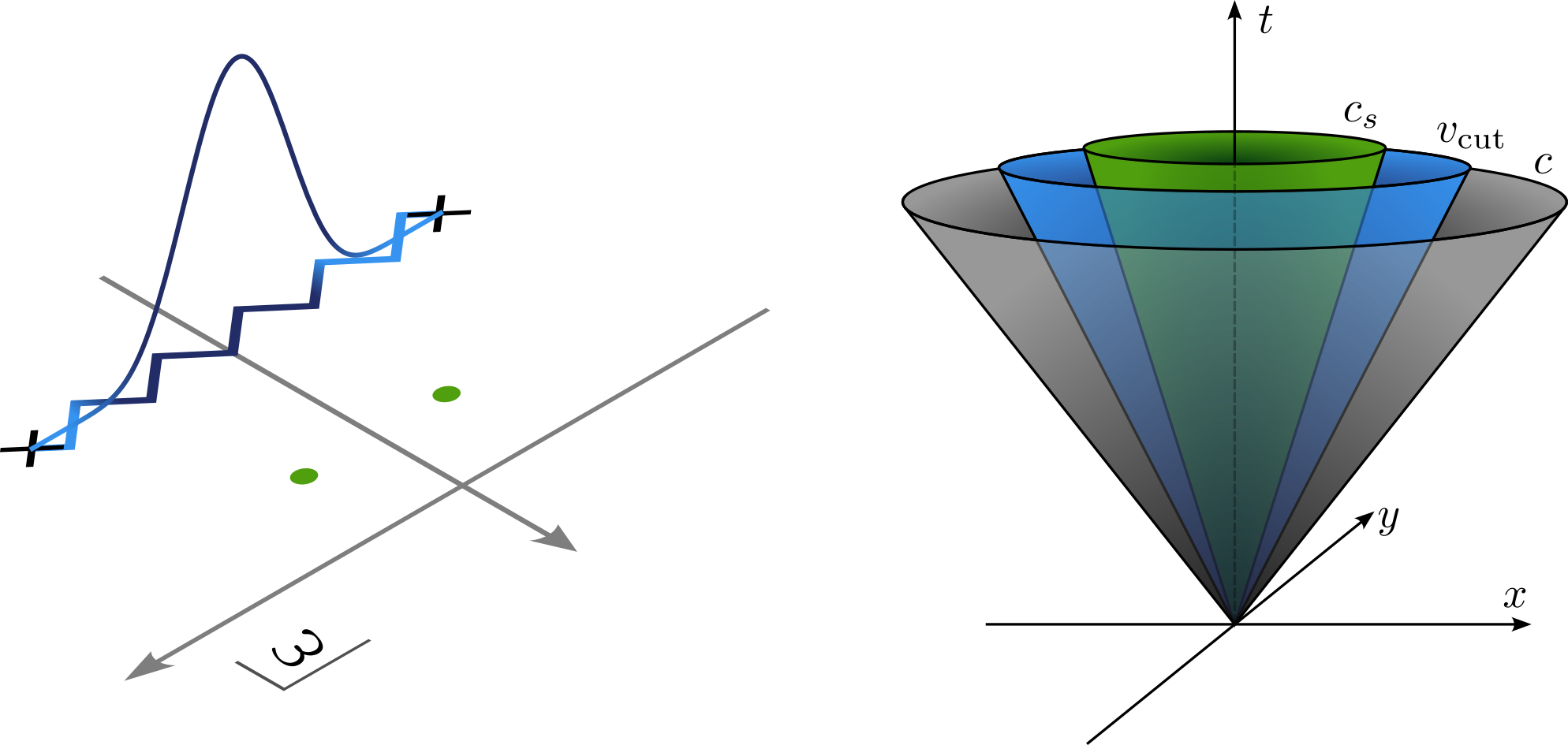}
    \caption{The structure of the RTA energy-energy correlator $G^{00,00}_{TT}$ in the complex frequency plane (left) can be related to the causal structure (right). The end points of the cut, denoted by crosses at $\omega= \pm ck-i/\tau_R$, encode the maximal propagation speed. This sets the causal lightcone, depicted in grey on the right figure. Furthermore, the discontinuity profile of the correlator along the cut, provides us with an effective velocity. The discontinuity becomes substantial at $\omega = \pm v_\mathrm{cut}k -i/\tau_R$. We interpret this $v_\mathrm{cut}$ as typical finite ballistic propagation, which sets an `effective' lightcone, denoted with blue on the right figure. Finally, the green dots denote the poles of $G^{00,00}_{TT}$, which capture the collective hydrodynamic modes, including the speed of sound. }
    \label{fig:lightcone}
\end{figure}

In the present work, we provide a complete understanding of the analytic structure of correlators in the RTA for a massive gas, both in terms of transport as well as providing a link between the complex structure of the correlator and causality. Previous works, e.g. \cite{Romatschke:2011qp}, have computed the thermodynamic quantities and some transport coefficients in the massive gas. We add to these results by providing closed form expressions for transport coefficients (in all spatial dimensions $d>1$, see \Cref{app:diffusion-arb}), as well as computing the thermoelectric coefficients for a massive gas.
We would like to point out some inaccuracies in \cite{Hataei:2025mqf}, which among other claims, indicated that the diffusive mode in RTA kinetic theory becomes unstable for a certain value of mass. We show that this is not the case for any $d>1$. Furthermore, we demonstrate that the correlators in the massive case possess a cut in the complex $\omega$ plane along which the correlator has a discontinuity. We find this cut to be undeformable and therefore refrain from using the term `branch cut', opting to refer to it as simply the `cut', unlike the proposed recent findings of \cite{Lin:2025ehr}, where they propose an undeformable branch cut arises as a `condensation' of continuously many branch points. We furthermore find that the discontinuity cut ends in what were branch points in the massless case $\omega = \pm k -i/\tau_R$ and provide a physical interpretation of the endpoints as well as the discontinuity profile along the cut. The essential picture is: 
\begin{itemize}[label=--]
    \item the location of the end points of the cut in complex frequency space corresponds to the fastest a signal can travel, i.e. the speed of light,
    \item the profile of the discontinuity along the cut provides a measure of the effective velocity of the particles,
    \item and the poles correspond to collective, i.e.~hydrodynamic,\footnote{As was the case in the $m=0$ case \cite{Bajec:2024jez}, we additionally find gapped, quasi--hydrodynamic poles \cite{Grozdanov:2018fic} in two spatial dimensions.} excitations.
\end{itemize}
We illustrate this by considering the energy-energy correlator for the massive gas in the complex frequency plane, shown in \Cref{fig:lightcone}. The analytic structure in the left panel is given by two hydrodynamic poles (shown in green) and a cut deeper in the complex plane, stretching between the two crosses. From the hydrodynamic modes it is straightforward to extract the speed of sound, which corresponds to the effective speed of collective excitations in this channel. The end points of the cut denote the maximal speed of propagation of particles in the present system.   Additionally, we show the discontinuity profile along the cut. In the present paper, we study the properties of the discontinuity profile as a function of mass and show, amongst other results, that increasing mass leads to less support along the cut. We then propose a condition, \cref{eq:cut-criteria}, which defines the velocity, $v_{\rm cut}$, which encodes the effective lightcone of massive particle propagation. To the best of our knowledge, the extraction of physical information from the discontinuity profile of the retarded correlator is novel. 
It will be curious to understand how general causal arguments \cite{Heller:2022ejw,Heller:2023jtd,Gavassino:2023mad} are modified in the RTA and the presence of cuts.

We would like to stress that although the results of this work are based in RTA kinetic theory, the analytic structures found therein are ubiquitous, especially at weak coupling. For example, it is known that gluon propagators in free Yang-Mills theory exhibit a regular tower of branch cuts at finite temperature \cite{Hartnoll:2005ju} and that a single branch cut dominates the non-hydrodynamic response of massless $\phi^4$ scalar field theory in linearized kinetic theory \cite{Rocha:2024cge}. Moreover, it will be curious to see how the massive RTA case corresponds to massless CFT \cite{Casalderrey-Solana:2018rle,Grozdanov:2018gfx} and massive field theory results, see e.g. \cite{Gursoy:2007cb,Gursoy:2007er,Betzios:2017dol}.

The structure of the paper is as follows: in \Cref{sec:setup}, we provide the necessary machinery to determine the retarded correlators in massive RTA kinetic theory and provide an overview of the thermodynamic properties. In \Cref{sec:transport}, we compute the explicit mass-dependent thermal, electric and thermoelectric transport coefficients in RTA kinetic theory. In \Cref{sec:thecut}, we discuss the shape of the discontinuity along the cut due to the mass and provide a physical interpretation. In \Cref{app:diffusion-arb}, we provide the expression for the charge and shear diffusion and discuss the stability of the massive theory in arbitrary dimensions. In \Cref{app:discontinuity} we describe the procedure to determine the discontinuity profile with explicit illustrative examples.

\section{Set-up}\label{sec:setup}

In this section we provide the reader with a brief overview of RTA kinetic theory with a constant relaxation time. We will follow the discussion and conventions from \cite{Bajec:2024jez}. 

We begin by considering the RTA Boltzmann equation for the one particle distribution function $f=f(t,\textbf{x},\textbf{p})$
\begin{align}
p^\mu \partial_\mu  f +  \mathcal{F}^i \partial_{p^{i}}f=\frac{p_\mu u^\mu}{\tau_R}\left( f - f_{eq}\right),
\end{align} 
where $\mathcal{F}^i$ is the external force, e.g.~the electric and gravitational force, can be encoded via
\begin{align}
    \mathcal{F}^\mu = F^\mn p_\nu - \Gamma^\mu_{\ab} p^\alpha p^\beta.
\end{align}
Here $F^{\mn}[ A_\mu]$ is the field strength tensor and $\Gamma^\mu_{\ab}[ g_\mn]$ is the Christoffel symbol. Note that the four velocity, $u^\mu$, is timelike normalized $u^\mu u_\mu=-1$ and $\tau_R$ is the relaxation time which we take to be constant.
In writing the above, we have assumed the mass-shell condition $p_0^2=p^2+m^2.$ 
At present, we will work with an equilibrium distribution function given by the Maxwell-J{\" u}ttner distribution, $f_{\rm eq}=e^{(p^\mu u_\mu+\mu_0)/T_0}$.
Moments of the one-particle distribution function correspond to macroscopic quantities, such as the number current and energy momentum tensor
\begin{align}\label{eq:cons-operators}
     J^\mu = \intp \frac{p^\mu}{p^0}  f, \quad \quad  T^\mn  =\intp \frac{p^\mu p^\nu}{p^0} f,
\end{align}
where $\intp =  \int {d^3 p}\,(2\pi)^{-3}.$

We are interested in extracting the linear response of the RTA kinetic theory in the case of massive particles. To do this, we turn on an external gauge potential, $\delta A_\mu$, and/or an external metric, $\delta g_\mn$. The presence of external fields induces a change in the macroscopic quantities via
\begin{align}
    \mu(t,\textbf{x})&= \mu_0 + \delta \mu(t,\textbf{x}),\\
    T(t,\textbf{x})&=T_0+\delta T(t,\textbf{x}),\\
    u^\mu(t,\textbf{x})&= u^\mu_0+\delta u^\mu(t,\textbf{x}),
\end{align}
where $u^\mu_0=(1,0,0,0).$
This in turn shifts both distribution functions 
\begin{align}
f(t,\textbf{x},\textbf{p})&=f_0(\textbf{p}) +\delta f(t,\textbf{x},\textbf{p}),\\
  f_{\rm eq}(t,\textbf{x},\textbf{p})&=f_0(\textbf{p}) +\delta f_{\rm eq}(t,\textbf{x},\textbf{p}),
\end{align}
where $f_0=e^{-\frac{p^0-\mu_0}{T_0}}$ and
the change to the equilibrium distribution function is given by
\begin{align}
    \delta f_{\rm eq}=\frac{f_0}{T_0}\left[\delta \mu + \frac{\delta T}{T_0}(p^0-\mu_0)+ p_i \delta u^i\right].
\end{align}
We parameterize the four momentum via $\tilde{v}^\mu =p^\mu / p^0 = (1, \gam v^i)$, where 
\begin{align}
    \gam = \frac{\abs{\textbf{p}}}{p^0}
\end{align}
and $v^i v_i = 1$. The linearized Boltzmann equation around the flat background $\eta_{\mu\nu} = \text{diag}(-1,\textbf{1})$ and at finite density $A^{(0)}_\mu = (\mu_0,\textbf{0})$ is then given by:
\begin{align}\label{eq:BoltzmannLinearized}
    p^\mu \partial_\mu \delta f + \mathcal{F}^i \frac{\partial f_0}{\partial p^i} = \frac{p\cdot u}{\tau_R}\left(\delta f - \delta f_\mathrm{eq}\right).
\end{align}
Note that the zeroth order solution is given by the equilibrium distribution $f_0=f_{\text{eq}}$.
 The solution of \eqref{eq:BoltzmannLinearized} in Fourier space\footnote{Our convention for the Fourier transform is $f(\omega,\textbf{k})=\int_{-\infty}^\infty dt \int d^3 x \, e^{i \omega t - i \textbf{k}\cdot \textbf{x}}f(t,\textbf{x})$.} is
\begin{align}\label{eq:gen-sol}
    \delta f = \frac{f_0}{T_0} \frac{ \gam  E^i v_i  -   \Gamma^0_\mn p^0 \tilde{v}^\mu \tilde{v}^\nu+\left(\delta\mu + p^i \delta u_i + \frac{\delta T}{T_0} (p^0-\mu_0)\right)/\tau_R }{  -i\omega + i \gam\,\textbf{k}\cdot \textbf{v} +1/\tau_R},
\end{align}
where $E^i=F^{0i}=i k^i A_0+i\omega A^i$ is the electric field. In the following computations, we will without loss of generality assume that perturbations are aligned with the $z$ direction, i.e.~that $\textbf{k}$ points along the $z$ direction.
It is then straightforward to compute changes to the conserved operators, namely the number current and the energy momentum tensor from \cref{eq:cons-operators}.
Additionally, requiring the conservation of the respective currents leads to the RTA matching conditions 
\begin{align}\label{eq:matching}
   \intp \frac{p\cdot u}{p^0}
   \left( f- f_{\rm eq}\right)&=0 , \qquad
   \intp p^\mu\frac{p\cdot u}{p^0}\left( f- f_{\rm eq}\right)=0.
\end{align}

To compute correlators, we will use the variational approach (see e.g. \cite{Romatschke:2015gic,Bajec:2024jez}), where we expand the conserved operators in powers of the external perturbations. Explicitly,
\begin{align}
        J^\mu&=J^\mu_0-G^{\mu,\nu}_{JJ}\delta A_\nu
        -\frac{1}{2}G^{\mu,\ab}_{JT}\delta g_{\ab}
        +\ldots\\
    T^\mn&=T^\mn_0-\frac{1}{2}G^{\mn,\ab}_{TT}\delta g_{\ab}
    -G^{\mn,\alpha}_{TJ}\delta A_{\alpha}
    +\ldots 
\end{align}
which leads to the definition of the correlators
\begin{align}\label{correlators}
G_{JJ}^{\mu,\nu}&= -\frac{\delta J^\mu}{\delta A_\nu}, \quad
G_{TJ}^{\mn,\alpha}= -\frac{\delta T^\mn}{\delta A_\alpha}, \quad G_{JT}^{\mu,\alpha\beta} = - 2 \frac{\delta J^\mu}{\delta g_{\ab}}, \quad
   G_{TT}^{\mn,\ab}= -2\frac{\delta T^\mn}{\delta g_\ab}.
\end{align}

\subsection{Thermodynamics at finite mass}\label{sec:thermo}

Here, we present the thermodynamics of a massive kinetic theory, restating known results in the literature, see e.g. \cite{Romatschke:2011qp}. We note that the static susceptibility is a function of the mass 
\begin{align}\label{eq:StaticSusc}
    \chi(m)= \frac{\partial n}{\partial \mu} = \frac{\partial}{\partial \mu}\int \frac{d^3 p}{(2\pi)^3} f_\text{eq} 
    &= T_0^2\frac{e^{\mu_0/T_0}}{2\pi^2 } \m^2  K_2\left(\m\right),
\end{align}
where $K_2(x)$ is the modified Bessel function of the second kind and $\m\equiv m/T_0$.
From \cref{eq:cons-operators}, the equilibrium number density $J^0=n_0$ is
\begin{align}\label{eq:n0}
    n_0 
    = \chi T_0.
\end{align}  
Similarly we can determine the equilibrium energy density, $T^{00}=\varepsilon,$ and pressure, $P_0 = T^{ii}$ (no sum), from \cref{eq:cons-operators} to be
\begin{align}\label{eq:energy0}
    \varepsilon_0 &= T_0^4 \frac{\m^2 
    }{2\pi^2}\left( \m K_1\left(\m \right)+3K_2\left(\m\right) \right),\\
    P_0 &= T_0^4 \frac{\m^2 
    }{2\pi^2}K_2\left(\m \right),\label{eq:pressure0}
\end{align}
in the absence of chemical potential.
In the limit $m\rightarrow0$ we recover the conformal equation of state, $P_0 = \varepsilon_0/3$ and $\varepsilon_0 \propto T_0^4$. 
From these expressions we can immediately determine the speed of sound squared at zero chemical potential, $c_s^2 = \partial P/\partial\varepsilon$, evaluated at local equilibrium
\begin{align}\label{eq:cs_EoS}
    c_s^2 = \frac{\m^2 K_3(\m)}{\left( \m^3 + 12 \m \right)K_0(\m) + \left(5\m^2 + 24\right)K_1(\m)},
\end{align}
which agrees identically with the result from \cite{Romatschke:2011qp}. For clarity, the small/large asymptotic behavior of the speed of sound is given by
    \begin{align}\label{eq:cs_Asymptotics}
        c_s&=
        \begin{cases}
            \frac{1}{\sqrt{3}}\left(1-\frac{1}{24}\m^2\right)+ \mathcal{O}\left(\m^4\right),
            & \m\ll 1,
            \\
         \frac{1}{\sqrt{\m}}\left(1-\frac{1}{4\m }\right)+\mathcal{O}\left({\m^{-5/2}}\right), 
        & \m\gg 1.
        \end{cases}
    \end{align}

\section{Transport properties at finite mass}\label{sec:transport}

In the present section, we provide the necessary steps to extract transport coefficients as a function of mass from the retarded correlators. 

\subsection{Charge transport}\label{sec:diffusion}

Here, we study the response in presence of an external electric field only, which induces a change in the chemical potential. In other words we turn off the external metric perturbations $\delta g_\mn = 0$, which in turn sets $\delta T = 0$ and $\delta u^i=0$. In this case, \eqref{eq:gen-sol} reduces to
\begin{align}
    \delta f = \frac{f_0}{T_0} \frac{\delta\mu/ \tau_R + \gam E^i v_i }{ -i\omega + i \gam\,\textbf{k}\cdot \textbf{v}+1/ \tau_R}.
\end{align}
Recalling that the change in the number density is related to the change in the chemical potential via the susceptibility, $\delta J^0=\delta n=\chi \delta \mu$,
a simple calculation leads to the density-density correlator:
\begin{align}\label{eq:corr00}
    G^{0,0}_{JJ} = \frac{i k  \chi \int_0^\infty \frac{dp}{2\pi^2}\,p^2 \frac{f_0}{T_0} I^{03}}{\chi - \int_0^\infty \frac{dp}{2\pi^2}\,p^2 \frac{f_0}{T_0}I^{00}},
\end{align}
where the static susceptibility $\chi$ is given in \eqref{eq:StaticSusc} and 
\begin{align}\label{eq:iab}
    I^{ab}(p)&= \int \frac{d\Omega}{4\pi}\frac{\tilde{v}^a \tilde{v}^b}{1 + \tau_R\left(-i\omega + i k\gam\cos\theta\right)},
\end{align}
where $\tilde{v}_a=(1,\gam v^i)$.  Explicitly, 
\begin{align}
    I^{00} &= 
    \frac{1}{2i\gam\,k\tau_R} L_\gam, \quad
    I^{03} =
    \frac{1 - i \omega \tau_R}{2 \gam \, k^2 \tau_R^2}L_\gam + \frac{1}{i\,k\tau_R},
\end{align}
where we defined the shorthand for the mass dependent logarithm
\begin{align}\label{eq:log}
    L_\gam=\log\left(\frac{\omega - \gam\, k + i/\tau_R}{\omega + \gam\, k + i/\tau_R}\right).
\end{align}

Therefore the pole of the correlator \cref{eq:corr00} is given implicitly by:
\begin{align}\label{eq:DiffusivePoleEq}
    \int_0^\infty d p \, p^2 \frac{f_0}{T_0} \left(1-I^{00}\right) = 0.
\end{align}
In the massless case, $I^{00}$ does not depend on $p$. Therefore the above integral equation is satisfied when
\begin{align}
    I^{00}\big\vert_{m=0} = 1 \Leftrightarrow \omega = \frac{i}{\tau_R}\left( k\tau_R \cot(k\tau_R)-1 \right),
\end{align}
where we used $\gam\big\vert_{m=0} = 1$. Expanding for small $k$ to quadratic order leads to the value of the charge diffusion constant $D = \tau_R/3$. This agrees with the massless result \cite{Romatschke:2015gic}.
    
    Returning to the massive case, we can expand \eqref{eq:DiffusivePoleEq} in small $\omega$ and $k$ up to quadratic order, which leads to
    \begin{equation}
    \int_0^\infty dp\,p^2
    f_0(p)
    \left(\gam ^2 k^2 \tau_R +3 \omega  (\tau_R  \omega -i) \right)
    = 0.
    \end{equation}
    The above equation can be solved analytically after integration to obtain:
    \begin{align}
        \omega(k)=-i D(m) k^2 + \mathcal{O}(k^4),
    \end{align}
where
\begin{align}\label{eq:FullChargeDiffusion}
        \frac{D(m)}{\tau_R} =
        \frac{1}{3}+ \m \frac{ G_{1,3}^{2,1}\left(\frac{\m^2}{4}\Bigg\vert
\begin{array}{c}
 1 \\
 -\frac{1}{2},\frac{1}{2},0 \\
\end{array}
\right)}{12  K_2\left(\m\right)}+\m\frac{\pi   }{6 
   K_2\left(\m\right)},
    \end{align}
    where $G^{2,1}_{1,3}$ is the Meijer $G$--function (see \cite{Gradshteyn:1702455} for definitions). In the limit $\m\rightarrow 0$ the above equation reduces to the known result $\omega(k) = - i \tau_R k^2 /3$ \cite{Romatschke:2015gic}. One can obtain the analytic expressions for the two limits $\m\rightarrow 0$ and $\m\rightarrow \infty$, namely
\begin{align}\label{eq:Diffusion_asymp}
        \frac{D}{\tau_R} =\begin{cases}
            \frac{1}{3} - \frac{1}{6} \m^2+ \mathcal{O}(\m^3) ,
            & \m\ll 1,\\
            \frac{1}{\m} + \mathcal{O}\left(\m^{-2}\right) ,
            & \m\gg 1.
        \end{cases} 
    \end{align}

We would like to stress that the diffusion coefficient remains positive for all values of the mass. There is no instability arising from a negative value of mass, as claimed by \cite{Hataei:2025mqf}. We believe that this claim came from an erroneous extrapolation of the small mass result. Furthermore, we can determine the diffusion coefficient in arbitrary dimension, the details of which we present in \Cref{app:diffusion-arb}, where we also comment on the absence of instability in arbitrary spatial dimensions $d>1$.

\begin{figure}
    \centering
    \includegraphics[width=0.49\textwidth]{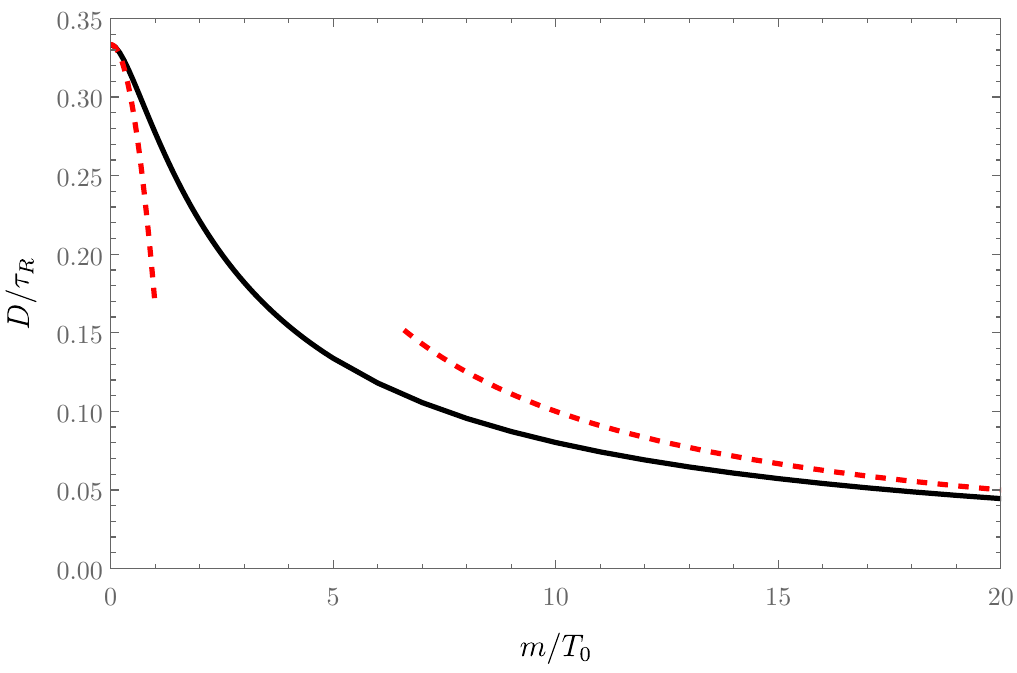}
    \includegraphics[width=0.49\linewidth]{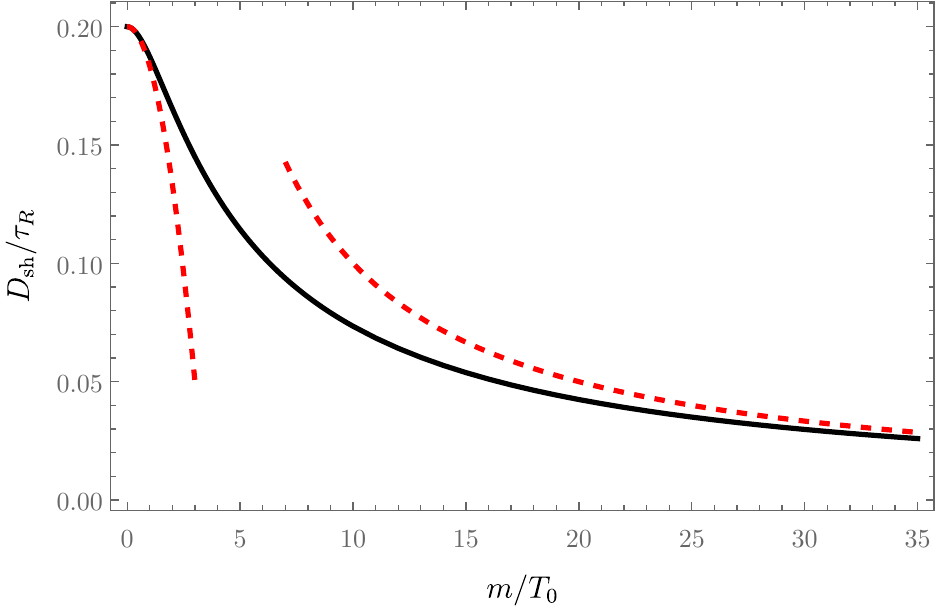}
    \caption{Left: The diffusion coefficient $D(m)$, as a function of mass at $T_0\tau_R=1$. The black line  denotes the full solution \eqref{eq:FullChargeDiffusion}, while the two red dashed lines denote the asymptotic behavior \eqref{eq:Diffusion_asymp}.
    Right: The shear diffusion coefficient $\Dsh(m) = \eta/(\varepsilon_0+P_0)$, as a function of mass in units where $\tau_R T_0=1$. The black line denotes the full solution \eqref{eq:FullShearDiffusion}, while the red dashed lines denote the asymptotic behavior \eqref{eq:ShearDiff_Asymptotics}.\label{fig:diffusion} 
    }
\end{figure}

\subsection{Momentum transport}

In the present section, we turn to momentum transport, which is mediated via a background metric perturbation. The background metric perturbation induces a change in the temperature and the four velocity. We turn off the background electric field and consequently, $\delta \mu=0.$ Then \cref{eq:gen-sol} reads
\begin{align}
    \delta f = \frac{f_0}{T_0} \frac{ \left( p^i \delta u_i + p^0\frac{\delta T}{T_0} \right)/\tau_R -   \Gamma^0_\mn p^0 \tilde{v}^\mu \tilde{v}^\nu }{  -i\omega + i \gam\,\textbf{k}\cdot \textbf{v} +1/\tau_R}.
\end{align}

\subsubsection{Shear channel}

We perform a similar derivation of the diffusion constant equation as in the charge diffusion case. The simplest case is the shear diffusion channel, schematically denoted with $\delta T^{0\perp}$, where $\perp$ is one of the transverse components to our chosen axis aligned with the wave vector $\textbf{k}$. For concreteness, we fix the transverse direction to $\perp = y$; other choices are physically equivalent. Because of the decoupled nature of the spin $1$ channel, we turn off sources that couple to $\delta T$ as well as $\delta u^j =0$ with $j\neq y$
. In this case, the self--consistency condition, $\delta T^{0y} = (\varepsilon_0 + P_0)\delta u^y$,  is trivially solved. A short calculation then shows:
\begin{align}
    \delta T^{0y} = \frac{-\tau_R \int_0^\infty \frac{dp}{2\pi^2}\,\frac{f_0}{T_0} p^2 (p^0)^2\gam\, \jsh}{1 - \frac{1}{\varepsilon_0+P_0}\int_0^\infty \frac{dp}{2\pi^2}\,\frac{f_0}{T_0} p^2 (p^0)^2
    \,I^{11}}
\end{align}
where $I^{ab}$ was defined in \eqref{eq:iab} and  
\begin{align}
        \jsh &\equiv \int\frac{d\Omega}{4\pi}\frac{v^y\Gamma^0_\mn\tilde{v}^\mu\tilde{v}^\nu}{1 + \tau_R\left(-i\omega + i k\gam\cos\theta\right)}.
\end{align}

The denominator structure encodes all the information about the collective modes in the spin $1$ channel. Using $\varepsilon_0 + P_0 = \frac{\partial T^{0y}}{\partial u^y} = \frac{1}{3}\int_0^\infty \frac{dp}{2\pi^2} \, \frac{f_0}{T_0}p^2 (p^0)^2\gam^2$, we obtain the integral equation describing shear diffusion:
\begin{align}
    \int_0^\infty \frac{dp}{2\pi^2}\,\frac{f_0}{T_0} p^2 (p^0)^2  \left(\frac{\gam^2}{3} - I^{11}\right) = 0.
\end{align}
The above equation can be generalized to arbitrary spatial dimensions, see \Cref{app:diffusion-arb}.
The integral can be solved analytically in the hydrodynamic limit by expanding the LHS up to cubic terms with respect to $k$ and $\omega$ in order to obtain the diffusive dispersion relation:
\begin{align}
    \omega(k) = -i D_\text{sh}(m) k^2 + \mathcal{O}(k^4),
\end{align}
where
\begin{align}\label{eq:FullShearDiffusion}
    \frac{D_\text{sh}}{\tau_R}=\frac{1}{5}+\m^2\frac{ G_{1,3}^{2,1}\left(\frac{\m^2}{4 }\Bigg\vert
\begin{array}{c}
 1 \\
 -\frac{3}{2},\frac{1}{2},0 \\
\end{array}
\right)}{40 K_3\left(\m\right)}-\m^2\frac{\pi}{30 
   K_3\left(\m\right)}.
\end{align}
The small and large mass limits of \eqref{eq:FullShearDiffusion} are given by
\begin{align}\label{eq:ShearDiff_Asymptotics}
    \frac{D_\text{sh}}{\tau_R} = \begin{cases}
        \frac{1}{5} - \frac{1}{60} \m^2 +\mathcal{O}\left(\m^4\right) & \m\ll 1,\\
        \frac{1}{\m} + \mathcal{O}\left(\m^{-2}\right) & \m\gg 1.
    \end{cases}
\end{align}
We show this result in the left panel of \Cref{fig:diffusion}, where we note that shear diffusion is positive for all values of the mass for any number of spatial dimensions $d>1$ (see \Cref{app:diffusion-arb}). This is in decided contrast to the results of \cite{Hataei:2025mqf}, who predicted that the shear diffusion should become negative for a certain value of $m/T_0.$ Comparing with results obtained from hydrodynamics \cite{Kovtun:2012rj}, we can identify $\Dsh = \eta/(\varepsilon_0+P_0)$. An independent calculation will cross check and confirm this result in \Cref{sec:OtherTransportCoeff}.

\subsubsection{Sound channel}

We now consider the longitudinal channel, i.e.~the sound. Here we only turn on sources that induce $\delta T$ and $\delta u^z$, since $k^i \propto \delta^i_z$. In order to solve the self--consistency condition and obtain the analytic structure of the correlators in this channel, one must solve two coupled linear equations for the perturbations $\delta T$ and $\delta u^z$ or equivalently for $\delta T^{00}$ and $\delta T^{03}$ using $\delta T^{00} = \frac{\partial \varepsilon_0}{\partial T_0} \delta T$ and $\delta T^{0z} = (\varepsilon_0+P_0)\delta u^z$:
\begin{align}
    &\delta T^{00}= \frac{\partial\varepsilon_0}{\partial T_0}\delta T = \Ip_{00}\frac{\delta T}{T_0} + \Ip_{03} \delta u^z + J_\text{s}^0 \nonumber,\\
    &\delta T^{0z}= (\varepsilon_0 + P_0)\delta u^z = \Ip_{03}\frac{\delta T}{T_0} + \Ip_{33} \delta u^z + J_\text{s}^z,\label{eq:SoundSystem}
\end{align}
where \begin{align}
    \mathcal{I}_{ab}&\equiv \int_0^\infty\frac{dp}{2\pi^2} \frac{f_0}{T_0} p^2  (p^0)^2 I_{ab},\\
    J^{a}_\text{s} &\equiv \intp\frac{1}{p^0} \frac{f_0}{T_0}p^a p^0 \frac{\tau_R \Gamma^0_\mn \tilde{v}^\mu \tilde{v}^\nu}{1 + \tau_R(-i\omega + i k \gam \tau_R\cos\theta)},\quad a=0,z.
\end{align}
For the sake of completeness we explicitly write out the solution to the above system from which one can deduce the sound channel correlators:
\begin{align}
    &\delta T^{00} =  \frac{\frac{\partial \varepsilon_0}{\partial T_0} T_0 \left( \Ip_{33} J_\text{s}^0 - \Ip_{03} J_\text{s}^z - (\varepsilon_0+P_0)J_\text{s}^0\right)}{(\Ip_{03})^2 - \left( \Ip_{00} - \frac{\partial \varepsilon_0}{\partial T_0} T_0  \right) \left( \Ip_{33}-\varepsilon_0-P_0 \right) },\label{eq:deltaT00}\\
    &\delta T^{0z} =-\frac{(\varepsilon_0+P_0)\left( \Ip_{03} J_\text{s}^0 - \Ip_{00} J_\text{s}^0 +\frac{\partial \varepsilon_0}{\partial T_0} T_0 J_\text{s}^z\right)}{(\Ip_{03})^2 - \left( \Ip_{00} - \frac{\partial \varepsilon_0}{\partial T_0} T_0  \right) \left( \Ip_{33}-\varepsilon_0-P_0 \right) }.
\end{align}

To extract the information about the collective excitations in this channel it is enough to study the determinant of the linear system \eqref{eq:SoundSystem}, or rather the denominator in \eqref{eq:deltaT00}. In the hydrodynamic limit, where we expand in collective powers of $\omega$ and $k$, we can extract the dispersion relation
\begin{align}\label{eq:SoundMode}
    \omega(k) = a_0 \pm a_1 k - i \Gamma_s k^2 +\mathcal{O}(k^3).
\end{align}
One can again use insight from hydrodynamics \cite{Kovtun:2012rj} to identify 
\begin{align}
\Gamma_s = \frac{(2-2/d)\eta + \zeta}{\varepsilon_0+P_0} = \frac{4}{3}\Dsh + \frac{\zeta}{\varepsilon_0+P_0},
\end{align}
where $\zeta$ is the bulk viscosity.\footnote{Using the notation of \cite{Kovtun:2012rj} we have $\Gamma_s = \frac{1}{2}\gam_s$.}

The coefficients of the dispersion relation \eqref{eq:SoundMode} are  given by
\begin{align}
    &a_0 = 0,\\
    &a_1 = c_s,\\
    \label{eq:bulk}
    \frac{\zeta}{\varepsilon_0+P_0} &=\frac{\tau_R}{72 \m^2 K_3} \Big(3 \m^4 G_{1,3}^{2,1}\left(\frac{\m^2}{4}\Bigg\vert
\begin{array}{c}
 1 \\
 -\frac{3}{2},\frac{1}{2},0 \\
\end{array}
\right) \\
&-\frac{1}{\left(\left(\m^2+8\right) K_1+4
   \m K_0\right) \left(\m \left(\m^2+12\right) K_0+\left(5 \m^2+24\right)
   K_1\right)}\nonumber \\
   &\times 4 \Big(24 \left(\m^2+8\right)^2 \left(\m^2+12\right)
   K_1^3-36 \m^6 K_3^3\nonumber\\
   &+\pi  \left(5 \m^4+64 \m^2+192\right) \m^4 K_1^2-96
   \left(\m^2-24\right) \m^3 K_0^3\nonumber\\
   &+\m K_0 K_1 \left(\pi  \m^4
   \left(\m^4+40 \m^2+192\right)-6 \left(\m^6-40 \m^4-960 \m^2-4608\right)
   K_1\right)\nonumber\\
   &+4 K_0^2 \left(\pi  \m^6 \left(\m^2+12\right)-12 \m^2
   \left(\m^4-24 \m^2-288\right) K_1\right)\Big)\Big),\nonumber
\end{align}
where the Bessel functions depend only on $\m$ and we have omitted the arguments for brevity. We note that as expected, the coefficient $a_1$ is given by the speed of sound $c_s$, calculated from the equation of state \cref{eq:cs_EoS}.
Using the above and \cref{eq:FullShearDiffusion}, one can reconstruct the full attenuation $\Gamma_s$. Alternatively, the bulk viscosity can be extracted via the Kubo formula
\begin{align}
    \zeta&=-\lim_{\omega\rightarrow0}\lim_{k\rightarrow0} \frac{1}{9\omega}\Im G^{ii,jj}_{TT},
\end{align}
which we find agrees with \cref{eq:bulk}.
The modes are gapless and propagating with the speed of sound \eqref{eq:cs_EoS}, calculated from the thermodynamic properties of our system. The attenuation is non--negative for any mass $m$. 

As before, we compute the small and large mass expansion for the transport coefficients.
For the attenuation, $\Gamma_s$,  the asymptotics are given by
\begin{align}\label{eq:attenuationAsymptotics}
\Gamma_s&=
        \tau_R\begin{cases}
            \frac{2  }{15}\left(1-\frac{1}{12}\m^2+ \mathcal{O}\left(\m^3\right)\right),
            & \m\ll 1,
            \\
         \frac{1}{\m}+\mathcal{O}\left(\m^{-2}\right), 
        & \m\gg 1.
        \end{cases}
\end{align}
and the bulk viscosity $\zeta$ 
\begin{align}\label{eq:BulkAsymptotics}
    \frac{\zeta}{\varepsilon_0 + P_0} = \tau_R\begin{cases}
            \frac{5}{432}\m^4 - \frac{\pi}{144}\m^5+ \mathcal{O}\left( \m^6 \right),
            & \m\ll 1,
            \\
         \frac{2}{3}\m^{-1}- \frac{16}{3}\m^{-2}+\mathcal{O}\left(\m^{-3}\right), 
        & \m\gg 1.
        \end{cases}
\end{align}

\begin{figure}
    \includegraphics[width=0.49\linewidth]{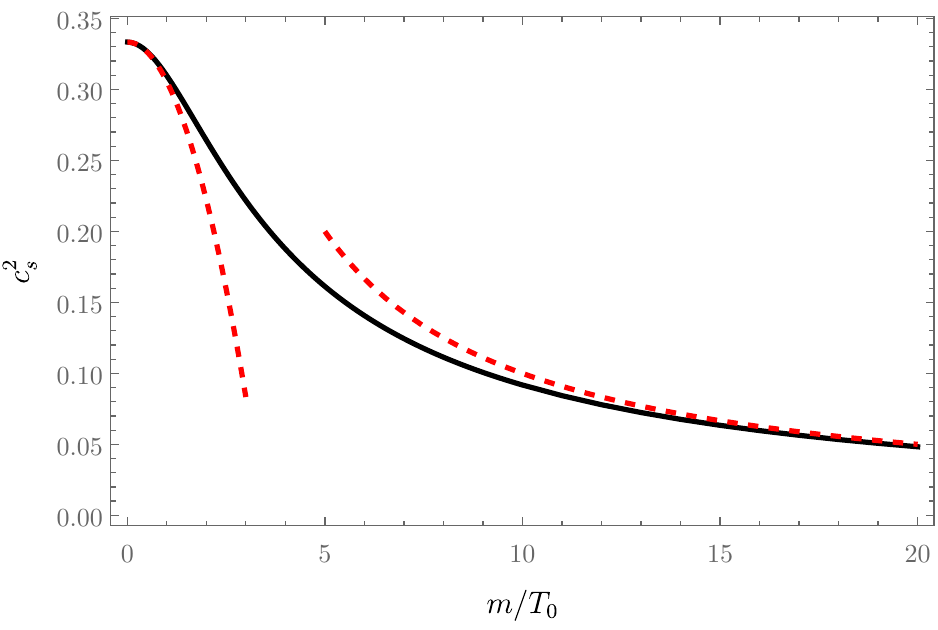}
    \includegraphics[width=0.49\linewidth]{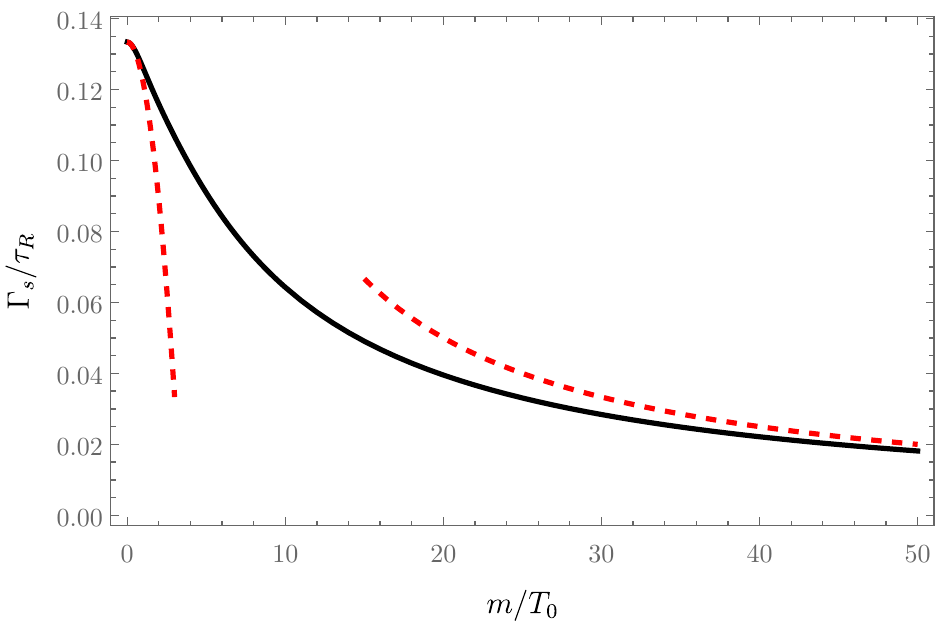}
    \includegraphics[width=0.49\linewidth]{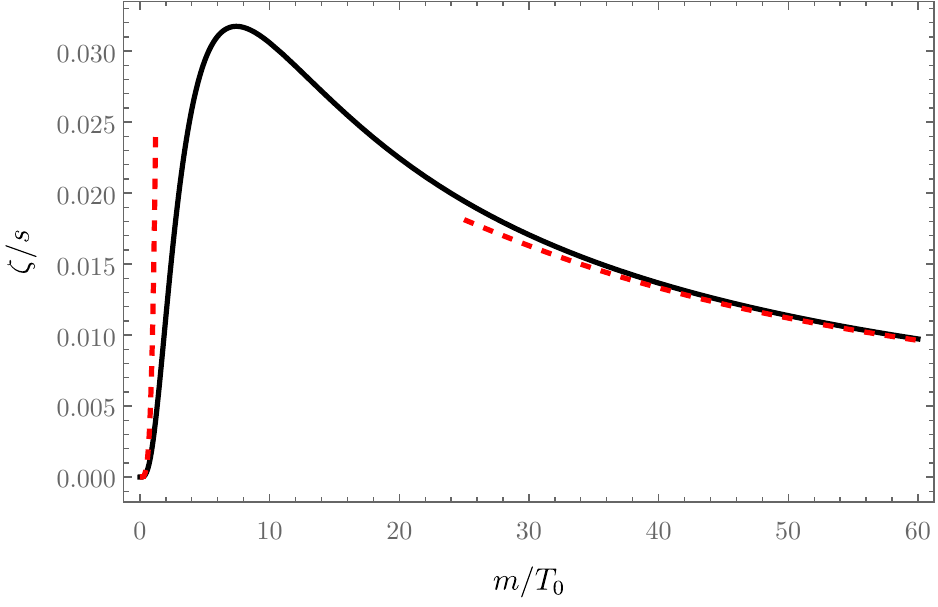}
    \includegraphics[width=0.49\linewidth]{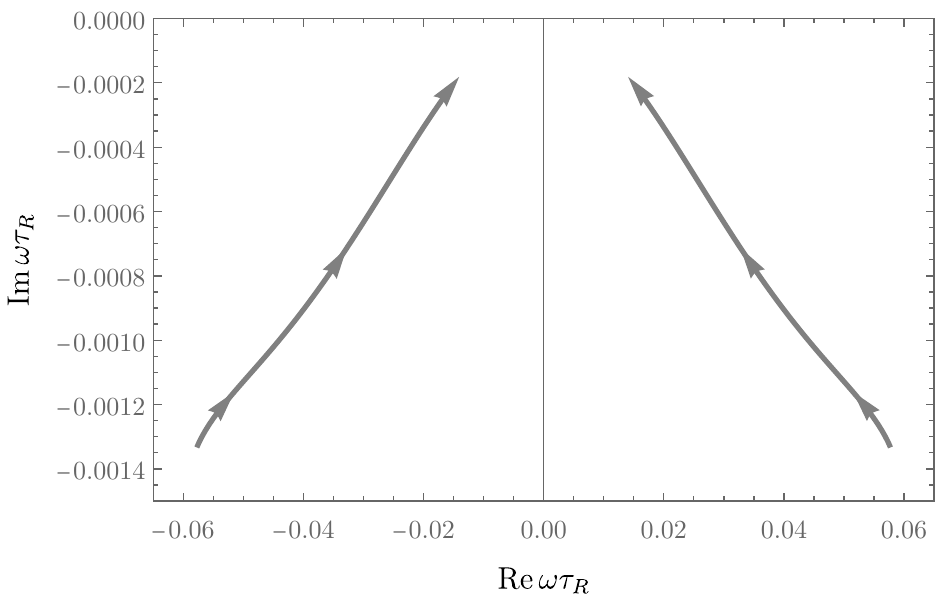}
    \caption{Top left: speed of sound squared with asymptotics given by \eqref{eq:cs_Asymptotics}. Top right: Dimensionless combination of attenuation with asymptotics given by \cref{eq:attenuationAsymptotics}.
    Bottom left: The dimensionless combination of bulk viscosity $\zeta$ over the entropy density $s$ as a function of mass in units where the relaxation time is $T_0\tau_R =1$. The red dashed lines corresponds to the analytic asymptotics given by \cref{eq:BulkAsymptotics}.
    Bottom right: Mass dependence of the sound pole in the  $G^{00,00}_{TT}$ correlator in the complex $\omega\tau_R$ plane with $k\tau_R=0.1$ as a function of $m/T_0.$
    }
\end{figure}

\subsubsection{Tensor channel 
}\label{sec:OtherTransportCoeff}

To determine the shear viscosity, it is enough to turn on just the metric perturbation, $\delta g_{xy}(\omega,k)$. This does not induce any change in the macroscopic variables ($\delta T$, $\delta u^i$), so there is no need to solve the matching conditions. The correlator is given by the following integral 
\begin{align}
    G^{xy,xy}&=-\frac{\delta}{\delta g_{xy}}\int \frac{d^3p}{(2\pi)^3} \frac{p^x p^y}{p^0}\delta f,\\
    &=-\int \frac{d^3p}{(2\pi)^3} \frac{p^x p^y}{p^0}\frac{\frac{\delta}{\delta g_{xy}}\Gamma^0_{\ab}\frac{p^\alpha p^\beta}{p^0}}{-i\omega +i\gam k \cos\theta+1/\tau_R}\frac{f_0}{T_0},\\
    &=-\int \frac{d^3p}{(2\pi)^3} \left(\frac{p^x p^y}{p^0}\right)^2\frac{2
    }{-i\omega +i\gam k \cos\theta+1/\tau_R}\frac{f_0}{T_0}.
\end{align}
Evaluating the integration over angles leads to
\begin{align}
    G^{xy,xy}=-\frac{\omega}{96 \pi ^2  k^5
   \tau_R^4} \int_0^\infty dp \,
   p (p^0)^3
   &\Big[
   i \gam  k \tau_R(1-i \tau_R \omega )\left(10  \gam ^2
   k^2 \tau_R^2 
   +6 
   (1-i \tau_R \omega )^2\right)\nonumber\\
   &-3
   \left(\gam ^2 k^2 \tau_R^2+(1-i \tau_R
   \omega )^2\right)^2 L_\gam\Big],
\end{align}
where $L_\gam$ was defined in \cref{eq:log}. It is more useful to change integration variable $p\rightarrow p^0=\sqrt{p^2+m^2}$. Expanding in small $k$ and using the Kubo formula, we then determine the shear viscosity via
\begin{align}
    \eta&=-\lim_{\omega\rightarrow0}\lim_{k\rightarrow0} \frac{1}{\omega}\Im G^{xy,xy}_{TT},\\
    &=\frac{\tau_R}{30 \pi ^2}\int_m^\infty dx \,x^{-1}\left(x^2-m^2\right)^{5/2}
   e^{-\frac{x}{T_0}}.\label{eq:eta}
\end{align} Note that the logarithm, $L_\gam,$ does not contribute in this limit. We note that the behavior of the shear viscosity over the entropy density, $s=(\varepsilon_0+P_0)/T_0$, where the energy density and pressure are given by \cref{eq:energy0,eq:pressure0}, is identical to the shear diffusion coefficient in \Cref{eq:FullShearDiffusion}. We note that the calculated transport coefficients, namely $\eta$ and $\zeta,$ agree with the results obtained using the Chapman-Enskog expansion presented in \cite{Hattori:2022hyo}.\footnote{Note that similar results for the bulk viscosity in the literature can be found, e.g.~\cite{ckbook}, which however vary slightly due to a different normalization convention of the distribution function.}

Similar to \cite{Romatschke:2015gic}, we can consider higher order transport coefficients, such as $\kappa$ and $\tau_\pi$, which we find to be independent of mass
\begin{align}
    \tau_\pi=\tau_R, \quad \kappa =0.
\end{align}

\subsection{Thermoelectric transport}
Finally, we turn our attention to the thermoelectric coefficients in the massive case, following the conventions of \cite{Bajec:2024jez}. The thermoelectric effect is determined by simultaneously turning on an external gauge field and metric perturbation. This means that we use the entire \cref{eq:gen-sol}, unlike in the previous subsections. The matching conditions \cref{eq:matching} relate the hydrodynamic fields  $(\delta n, \delta \varepsilon, \delta u^i)$ to the conserved operators $(\delta J^0, \delta T^{00},\delta T^{0i})$ via 
\begin{align}
    \delta J^0=\delta n, \quad \delta T^{00}=\delta \varepsilon, \quad \delta T^{0i}=(\varepsilon_0+P_0)\delta u^i.
\end{align} The hydrodynamic fields need to be in turn related to the temperature and chemical potential via \cite{Kovtun:2012rj}
\begin{align}
    \delta n &= \frac{\delta n_0}{\delta \mu_0} \delta \mu+ \frac{\delta n_0}{\delta T_0} \delta T,\\
    \delta \varepsilon &= \frac{\delta \varepsilon_0}{\delta \mu_0} \delta \mu+ \frac{\delta \varepsilon_0}{\delta T_0} \delta T,
\end{align}
where the equilibrium number and energy density are given in \cref{eq:n0} and \cref{eq:energy0}, respectively. 

The thermoelectric response, relating the change in the number flux and the heat current $\delta Q^i=\delta T^{0i}-\mu_0 \delta J^i$ due to an external electric field and temperature gradients, is captured via the following 
\begin{equation}\label{eq:TE_TransportMatrix}
    \begin{pmatrix}
        \delta J^i \\
        \delta Q^i
    \end{pmatrix} = \begin{pmatrix}
        \sigma^{ij} & T_0 \alpha^{ij} \\
        T_0 \tilde{\alpha}^{ij} & T_0 \bar{\kappa}^{ij}
    \end{pmatrix}
    \begin{pmatrix}
        E_j \\
        -\frac{1}{T_0}\nabla_j \delta T
    \end{pmatrix}.
\end{equation}
Recalling the relationship between the external electric and temperature fields to the gauge and metric perturbations \cite{Hartnoll:2007ih,Bajec:2024jez}, we have in Fourier space
\begin{align}
        E_j=i\omega \delta A_j + i\omega \mu_0 \delta g_{tj} ,\quad 
        -ik_j \frac{\delta T}{T_0}=i\omega \delta g_{tj}.
    \end{align}
This then leads directly to the Kubo formula for the thermoelectric transport coefficients
\begin{align}\label{kubo}
    \sigma^{ij}(\omega) &= -\frac{1}{i\omega}\lim_{k\rightarrow 0}\left(G^{ij}_{JJ}(\omega,k) - G^{ij}_{JJ}(0,k)\right),\\
    \tilde{\alpha}^{ij}(\omega) &=  -\frac{1}{i\omega T_0}\lim_{k\rightarrow0}\left(G^{ij}_{QJ}(\omega,k)-G^{ij}_{QJ}(0,k)\right),\label{kubo_alphat}\\
    \alpha^{ij}(\omega) &= -\frac{1}{i\omega T_0}\lim_{k\rightarrow0}\left(G^{ij}_{JQ}(\omega,k)-G^{ij}_{JQ}(0,k)\right),\label{kubo_alpha}\\
    \bar{\kappa}^{ij}(\omega) &= -\frac{1}{i\omega T_0}\lim_{k\rightarrow0}\left(G^{ij}_{QQ}(\omega,k) - G^{ij}_{QQ}(0,k)\right)\label{kubo_kappa},
\end{align}
where the correlators are defined in \eqref{correlators} and we use the shorthand
\begin{equation}
     \frac{\delta X^i}{\delta Q^j} = 2\frac{\delta X^i}{\delta g_{tj}} - \mu_0 \frac{\delta X^i}{\delta A_j}.
\end{equation}

For convenience, we express the thermoelectric coefficients in terms of the following integral, which we determined previously to be the diffusion coefficient \cref{eq:FullChargeDiffusion}
\begin{align}
    \frac{\tau_R}{6\pi^2T_0}\int_m^\infty dx\, e^{-x/T_0}\frac{(x^2-m^2)^{3/2}}{x}=\chi D.
\end{align}
The above equality is a consequence of $D$ satisfying \Cref{eq:ChargeDiff_dDim} in the $d=3$ case. Using \cref{kubo,kubo_kappa} and the fact that the correlators satisfy Ward identities
\begin{align}
    (\mu_0 \sigma+T_0 \alpha)i \omega&=-n_0,\\
    (\tilde{\kappa}+\mu_0 \alpha)i \omega&=-s_0,
\end{align} we observe that the thermoelectric transport can be parameterized in terms of one microscopic variable, namely the DC conductivity $\sigma_Q$, and thermodynamic quantities \cite{Hartnoll:2007ih,Hartnoll:2016apf}
\begin{align}
    \sigma&= \sigma_Q - \frac{1}{i\omega}\frac{n_0}{\varepsilon_0+P_0},\\
    \alpha=\tilde{\alpha}&=-\frac{\mu_0}{T_0}\sigma_Q- \frac{1}{i\omega T_0}\frac{n_0 s_0}{\varepsilon_0+P_0},\\
    \tilde{\kappa}&=\frac{\mu_0^2}{T_0}\sigma_Q- \frac{1}{i\omega T_0}\frac{T_0 s_0^2}{\varepsilon_0+P_0},
\end{align}
where $n_0$ is the equilibrium number density and $s_0 = (\varepsilon_0+P_0-\mu_0 n_0)/T_0$ is the equilibrium entropy density.
The frequency dependent terms correspond to the Drude pole.
We determine the zero frequency DC conductivity to be
\begin{align}\label{sigmaQ}
    \sigma_Q(m)= \chi D
    -\tau_R\frac{ n_0^2}{\varepsilon_0+P_0}.
\end{align}
We note that in the zero mass limit, this agrees with the result from \cite{Bajec:2024jez}
\begin{align}\label{sigmaQ0}
    \sigma_Q(m=0)=\frac{\tau_R }{12}\chi(m=0).
\end{align}
The small and large mass expansion of the zero frequency $\sigma_Q$ is given by
\begin{align}\label{eq:sigmaQ_Asymptotics}
    \frac{\sigma_Q}{\chi\tau_R} = \begin{cases}
        \frac{1}{12} - \frac{13 }{96}\m^2+\mathcal{O}\left(\m^3\right),& \m\ll 1,\\
        \frac{5}{3} \m^{-3}+\mathcal{O}\left(\m^{-4}\right),& \m\gg 1.
    \end{cases}
\end{align}

\begin{figure}[tbp]
    \centering
    \includegraphics[width=0.55\linewidth]{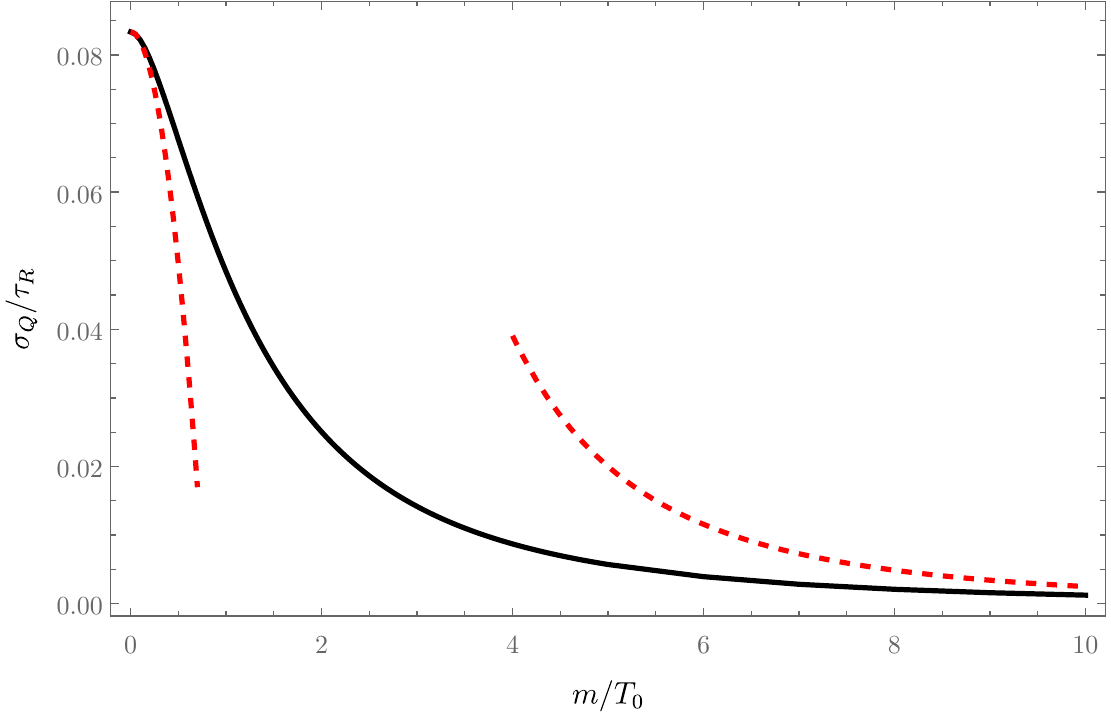}
    \caption{The microscopic contribution to the conductivity \cref{sigmaQ}, with asymptotics given by \Cref{eq:sigmaQ_Asymptotics}.}
    \label{fig:sigmaQ}
\end{figure}

\section{Non-hydrodynamic modes}\label{sec:thecut}

Having determined the transport coefficients as a function of mass, we move the discussion away from the poles of the correlators, which are captured by the underlying hydrodynamic theory, and focus our attention on the typical non-hydrodynamic signature of kinetic theory: the presence of cuts. For the present purposes, we will limit our attention to the correlator $G^{2,2}_{JJ}$, which is a particularly simple correlator due to the absence of matching conditions. Its nontrivial analytic structure is given solely by the RTA  cut. We note that similar considerations outlined below will qualitatively apply to all other correlators.

\begin{figure}
    \centering
    \includegraphics[width=\linewidth]{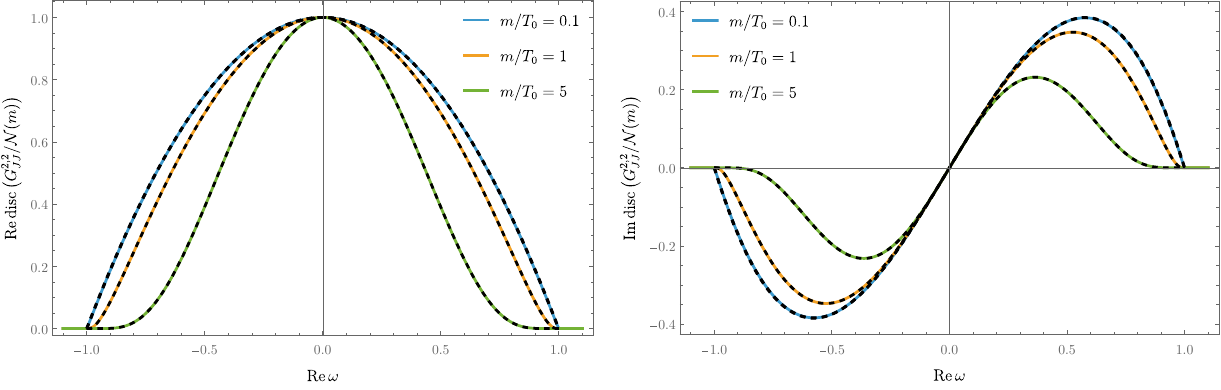}
    \caption{The normalized real and imaginary part of the discontinuity of the $G_{JJ}^{2,2}$ correlator for different values of $m$ in units where $T_0 = 1$ and $\tau_R = 1$ at fixed momentum $k \tau_R=1$. The dashed line corresponds to the analytic result \cref{eq:GJ22_DiscAnalytic}.}
    \label{fig:disc_gjj22}
\end{figure}

The correlator is given by
\begin{align}
    G^{2,2}_{JJ} &= -\frac{i \omega \tau_R}{4}\int_0^\infty \frac{dp}{2\pi^2} p^2\frac{f_0}{T_0} I_{11},\\
    &=\frac{\omega\tau_R}{8 \pi ^2 T_0 k^3
   \tau_R^3 }\int_0^\infty dp\, e^{\frac{-\sqrt{m^2+p^2}}{T_0}}\frac{  p^2    }{\gam }\left(2 \gam  k \tau_R (\tau_R \omega +i)- \left(\gam ^2 k^2 \tau_R^2-(\tau_R \omega +i)^2\right)L_\gam\right).\label{eq:gjj22}
\end{align}
We can extract the discontinuity profile by using the procedure described in \Cref{app:DiscontinuityProfile}. In this case, the discontinuity profile is
\begin{align}\label{eq:GJ22_DiscAnalytic}
    \mathrm{disc}\, G^{2,2}_{JJ}(\omega,k)=-\frac{ \left(k^2\tau_R^2-\left(\wRe\right)^2\tau_R^2\right) (1-i \,\wRe\tau_R)  \left(\frac{m}{T_0}
   \frac{k}{\sqrt{k^2-\wRe^2}}+1\right)}{2 \pi  k^3 \tau_R^3}e^{-\frac{m}{T_0}k \left({k^2-\left(\wRe\right)^2}\right)^{-1/2}},
\end{align}
where $\wRe$ denotes the real part of $\omega$ when positioned on the cut, i.e. $\omega = \wRe - i/\tau_R$. As such, the above result is valid for $\wRe\in[-k,k]$. 
Since the discontinuity profile is exponentially suppressed with mass $m$, we will normalize it by $\mathcal{N}(m) \equiv \mathrm{disc}\,G_{JJ}^{2,2}\big\vert_{\Re\omega = 0}$. In \Cref{fig:disc_gjj22}, we compare the above analytic result explicitly with numeric integration.

Taking a step back and leaving the integral over the polar angle $\theta$ unevaluated, we can rewrite our correlator as
\begin{align}
    G^{2,2}_{JJ} =-i\omega\tau_R \int_{-1}^1 \frac{d\cos\theta}{4}\int_0^\infty \frac{dp}{2\pi^2}p^2\frac{f_0}{T_0}\gam^2\frac{1-\cos^2\theta}{1 -i\omega\tau_R + i k\tau_R \cos\theta\,\gam}.
\end{align}
Introducing a new variable $u= \gam = \sqrt{1-\frac{m^2}{x^2}}$ and denoting $y=\cos\theta$ for brevity:
\begin{align}
    G^{2,2}_{JJ}& = -\frac{1}{T_0}\frac{\omega}{ k }\int_{-1}^1\frac{dy}{4 y}\int_0^1 \frac{du}{2\pi^2}\frac{m^3 u^4}{(1-u^2)^{5/2}}\exp\left( -\frac{m}{T_0} \frac{1}{\sqrt{1-u^2}} \right) \frac{1-y^2}{u - \frac{i\omega \tau_R - 1}{i k \tau_R y}}\\
    &\equiv\int_{-1}^1 dy\, \Gy(y;\omega) .
\end{align}
Following the discussion from \Cref{app:SingularIntegrals}, we find that for fixed angle $y=y_0$, the integration over $p$ provides acut at
\begin{align}
    \omega = s\, ky_0 - i/\tau_R,\quad s\in[0,1],
\end{align}
with the discontinuity profile given by
\begin{align}\label{eq:Energydisc-prof-22}
\mathrm{disc}\,\Gy(y_0;\omega)\bigg\vert_{\omega = s\,ky_0 - i/\tau_R} = \frac{m^3 s^4 \left(y^2_0-1\right) e^{-\frac{m}{T_0}\frac{1}{\sqrt{1-s^2} }} (1+i k s \tau_R y_0)}{4 \pi  k
   \left(1-s^2\right)^{5/2} T_0 \tau_R y_0}.
\end{align}
This discontinuity arising from the momentum integration is consistent with the physical interpretation of the branch cut from \cite{Kurkela:2017xis}, which we comment on in the next section. We note that studying a similar example in \Cref{fig:singularDiscontinuityExample_Check}, we see that taking the mass to zero corresponds to the support of the discontinuity decreasing everywhere except in the vicinity of the end points, where it becomes more strongly peaked. This can be interpreted that in the massless case, particles travel at a single speed (that of light), while in the massive case there is a distribution of velocities.

\subsection{Physical interpretation of analytic structure in the massive case}

Here, we interpret the behavior of the cut in the previous section for the massive gas. 
To build intuition, we begin with a non-interacting massless gas, restating the arguments in \cite{Kurkela:2017xis}. Particles are travelling in ballistic trajectories at the speed of light. Now, consider a plane wave perturbation in the sound channel. The response at some spacetime location $(t,\textbf{x})$ depends on the contribution over the sphere of radius $ct.$ Maximal signal, $\omega=k$,  will correspond to particles moving along the direction of the perturbation. Any other direction will have a reduced signal, $\omega=\textbf{v}\cdot \textbf{k}.$ This corresponds to a pole in the complex frequency plane. Integrating over all angles, the poles assemble into a logarithmic branch cut
\begin{align}\label{eq:massless-log}
    \int \frac{d\Omega}{4\pi} \frac{1}{-i\omega+i \textbf{v}\cdot \textbf{k}}=\frac{1}{2ik}\log{\frac{\omega-k}{\omega+k}}.
\end{align}
Including interactions via relaxation to equilibrium in the RTA leads to damping, which pushes the branch cut deeper into the lower half plane. Practically, this amounts to shifting $\omega\rightarrow \omega+i/\tau_R.$

In the massive non-interacting case, particles move at a distribution of speeds slower than the speed of light. As such, in addition to the angular integration, the situation in the complex plane is further complicated by needing to integrate over all momenta
\begin{align}\label{eq:MomentumIntegrationLog}
    \int dp\, p^2 f_0(p)\int \frac{d\Omega}{4\pi} \frac{1}{-i\omega+i \gam(p)\textbf{v}\cdot \textbf{k}}=\int dp\, p^2 f_0(p)\frac{1}{2i\gam(p)k}\log{\frac{\omega-\gam(p)k}{\omega+\gam(p)k}}.
\end{align}
We see that the branch cut structure arises both from the angular and momentum integration, which should be contrasted to the massless case where the continuous spectrum assembles from the continuous set of poles from solely the angular integration.  
The continuous string of signals coming from the momentum integration is a consequence of finite mass $m>0$ since now the speed of propagation of the individual particles between collisions is energy dependent (and not fixed to the speed of light as was the case in the massless theory), c.f. \Cref{eq:Energydisc-prof-22}.

While the explicit expression of the integral in \Cref{eq:MomentumIntegrationLog} is unattainable in closed form, one can  again follow the discussion from the previous sections as well as \Cref{app:DiscontinuityProfile} to extract the discontinuity profile. 
We see generically from \Cref{fig:disc_gjj22} that for increasing mass, the support is exponentially suppressed until some value of $v_{\rm cut}$.  
Taking \cref{eq:GJ22_DiscAnalytic} as an example, we determine the cut velocity via the following criterion
\begin{align}\label{eq:cut-criteria}
    \mathrm{disc}\, G^{2,2}_{JJ}(\omega_*,k)=\epsilon, 
\end{align}
from which we find $\omega_*=v_{\rm cut}k+\mathcal{O}(k^2)$. We extract the effective cut velocity via $v_{\rm cut}=d\omega/dk\vert_{k=0}.$ For a simpler, explicit and analytic example in the case of a logarithm, see \cref{eq:vcut}.

Interestingly, it seems that the structure of the discontinuity along the cut seems to encode causal information from the propagator. In the massless case, the branch points of the logarithm \cref{eq:massless-log} correspond to the maximum possible speed attainable by particles in the system, namely $\omega=\pm k$. 
It would seem natural that in the massive case a similar causal structure holds. Indeed, we find that the end points are the same as in the massless case. The major difference is that in the massive case, the argument of the logarithm is momentum dependent. As there is a distribution of particles of all different velocities, particles from different directions angular directions contribute with varying weight. The contribution of velocities $\gtrsim v_{\rm cut} $ is exponentially suppressed due to the distribution function. For velocities below $v_{\rm cut},$ more signal reaches the spacetime point.

\begin{figure}
    \centering
    \includegraphics[width=0.5\linewidth]{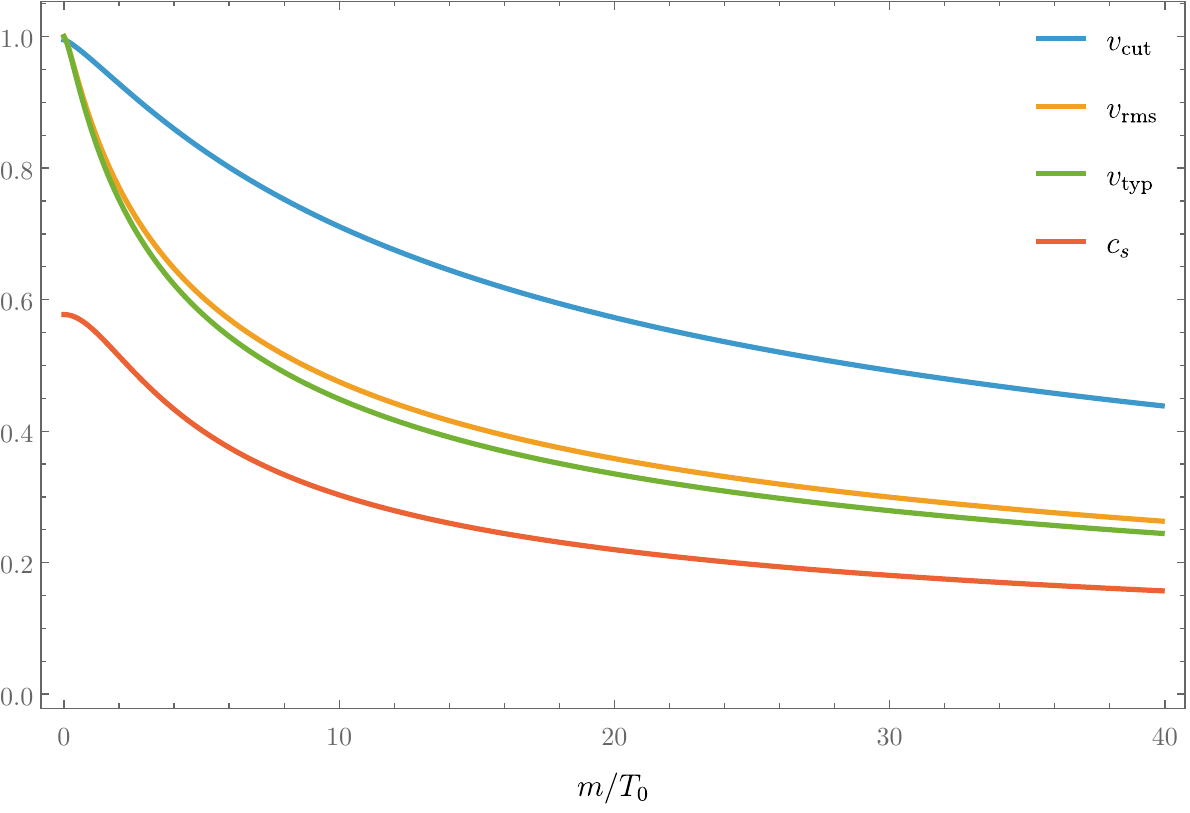}
    \caption{Comparison of the various speeds of the massive gas: the dynamical speed determined from the cut \cref{eq:cut-criteria}, $v_{\rm cut}$, the typical speed \cref{eq:vtyp}, $v_{\rm typ}$, rms speed of a gas \cref{eq:vrms}, $v_{\rm rms},$ and the speed of sound \eqref{eq:cs_EoS}.}
    \label{fig:speeds}
\end{figure}

We interpret this in terms of causal structure, with a picture in terms of lightcones found in \Cref{fig:lightcone}. The essential idea is that the location of the end points of the cut denote the maximum speed of propagation. Further, the profile of the discontinuity takes into account the distribution of velocities of the massive particles, setting roughly the size of the effective lightcone at the typical velocity \cref{eq:vtyp}. Finally, within the effective lightcone one finds the collective excitations of the system, such as that due to the speed of sound. To show that $v_{\rm cut}$ is distinct to the typical velocities computed in kinetic theory, in \Cref{fig:speeds} we compare the velocity determined from the cut, $v_{\rm cut}$, to the (normalized) typical and root-mean-square (rms) velocity in the absence of external fields 
\begin{align}
    v_{\rm typ} &\equiv \langle v\rangle=\frac{\int \frac{d^3 p}{(2\pi)^3 p^0  } \frac{p}{p^0}f_0   
    }{\int \frac{d^3 p}{(2\pi)^3 p^0  }f_0   }
    =  \frac{  \m^2\, \text{Ei}(-\m)+e^{-\m}\left(\m+1\right)}{\m \,K_1(\m)},\label{eq:vtyp}\\
        v_{\rm rms}^2 &\equiv \langle v^2\rangle=\frac{
        \int \frac{d^3 p}{(2\pi)^3 p^0  } \left(\frac{p}{p^0}\right)^2f_0
        }{\int \frac{d^3 p}{(2\pi)^3 p^0  }f_0 }
        \label{eq:vrms}
\end{align}
respectively, where $Ei(x)$ is the exponential integral, defined in \cref{eq:eiz}.

\begin{figure}
    \centering
    \includegraphics[width=0.35\linewidth]{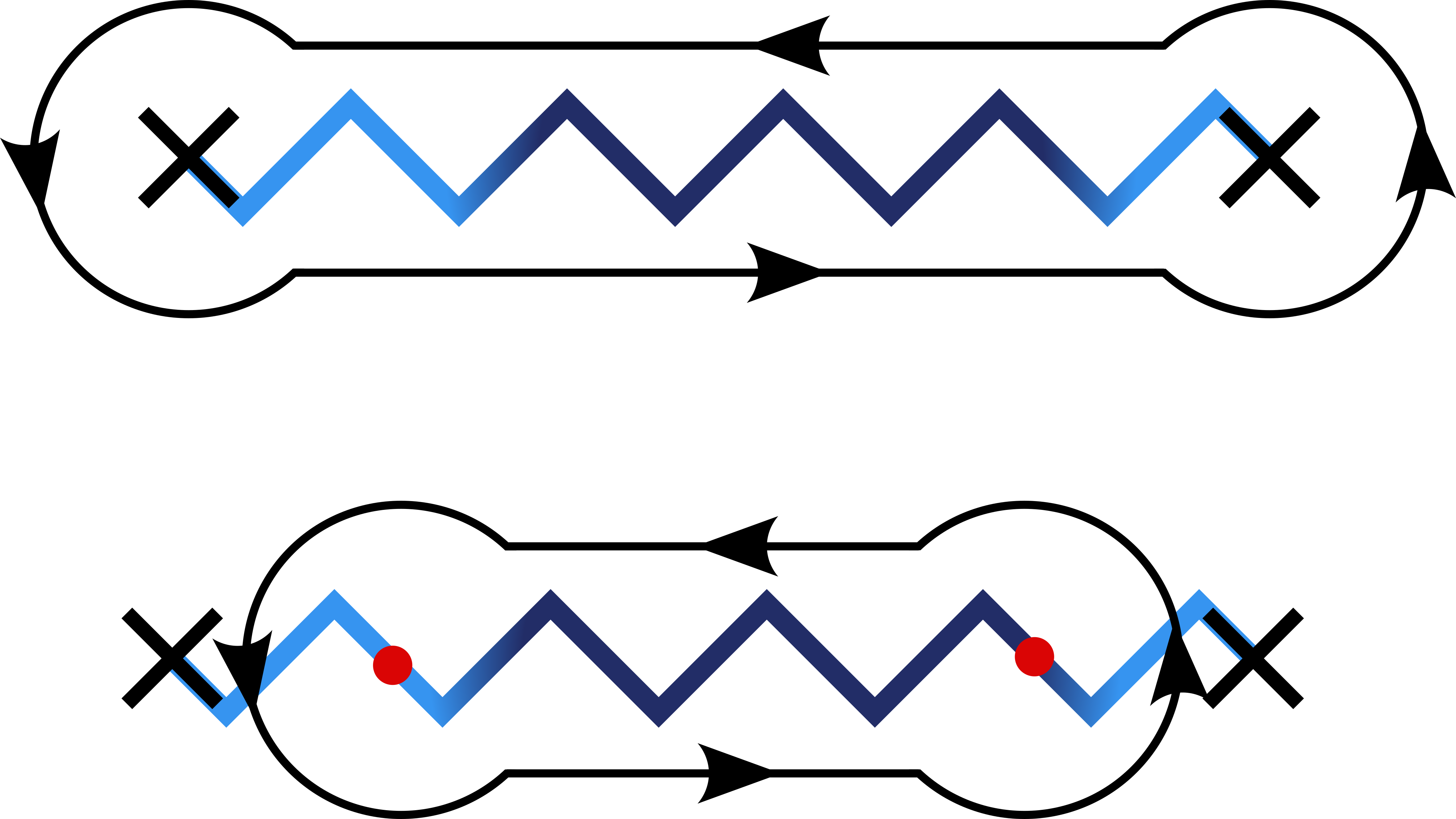}
    \caption{Top: usual dogbone contour integral used to evaluate integrals. Bottom: contour chosen following the $v_{\rm cut}$ criteria, \cref{eq:cut-criteria}.}
    \label{fig:dogbone}
\end{figure}

To gain some intuition in the real time domain, we turn to the Fourier transform of \cref{eq:gjj22}. We integrate in two different ways to better understand the pseudo-branch point determined by $v_{\rm cut}$: first, we take the standard dogbone contour integral over the whole cut for $\vert\omega\vert<k$, see the top of \Cref{fig:dogbone}, which provides the correct Fourier transform. Next, we evaluate the dogbone contour as if the end points were at the value of $\omega_*$ determined by \cref{eq:cut-criteria}. 

\begin{figure}
    \centering
    \includegraphics[width=\linewidth]{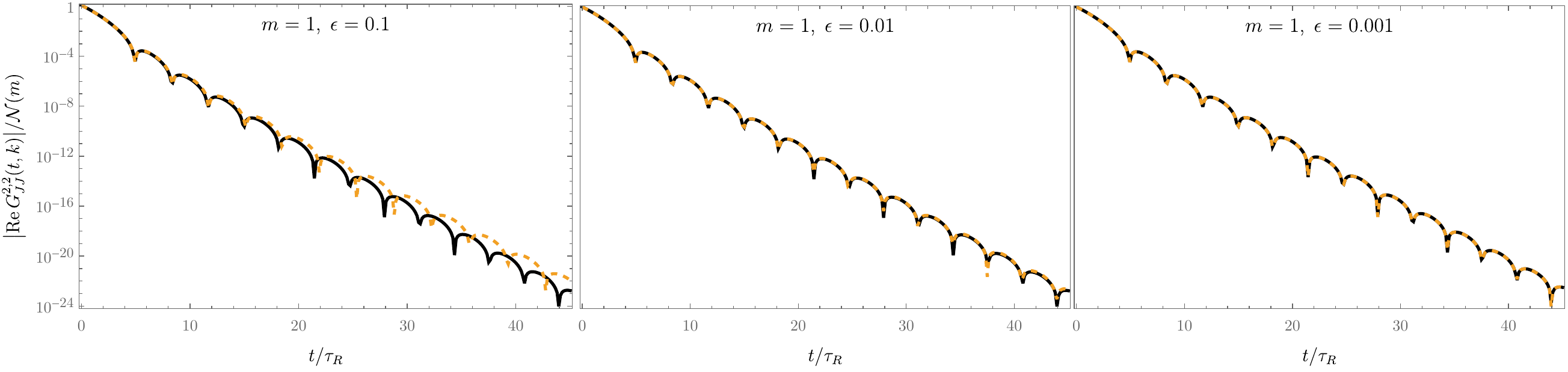}
    \includegraphics[width=\linewidth]{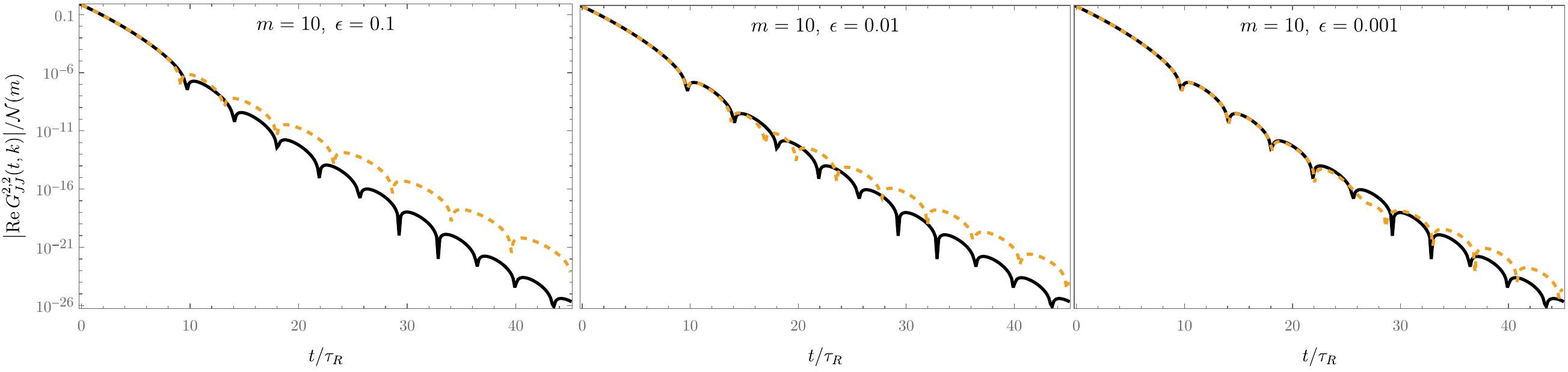}
    \caption{Real part of the Fourier transform of the $G^{2,2}_{JJ}$ normalized by its maximum value of the discontinuity for (top plots) $m/T_0=1$ and (bottom plots) $m/T_0=10$ for increasingly smaller values of \cref{eq:cut-criteria} (from left to right). 
    }
    \label{fig:FT_GJJ}
\end{figure}

We show the result of these different procedures in \Cref{fig:FT_GJJ} via the absolute value of the real part of the Fourier transform of \cref{eq:GJ22_DiscAnalytic} for two different masses $m/T_0=1,10$ and for a variety of cutoffs, defined in \eqref{eq:cut-criteria}. We see that there is excellent agreement for $\epsilon=10^{-3}$ over multiple oscillations and relaxation times for both masses. This agreement is even more impressive when one takes into account how much of the cut we are integrating over as shown in the bottom \Cref{fig:dogbone} -- for the bottom right panel of \Cref{fig:FT_GJJ}, we are integrating only around $\omega \sim \pm 0.8 k$, which represents approximately 20$\%$ less cut than taking the full contour. Finally, we note that we are only considering the non-hydrodynamic cut. As such, the disagreement at late times occurs both when the contribution of the non-hydrodynamic mode becomes small and in the regime when the hydrodynamic modes become dominant.

\subsection{Indeformability of the cut at finite mass}\label{sec:deform}
In this subsection, we discuss the freedom one has in usually choosing the branch cut. 
We begin by considering the branch cut in the correlators at zero mass. Such a structure stems from integrals over complex variables of the form
\begin{align}
    \int_{-1}^1 dz \frac{1}{1-i\omega\tau_R + i k\tau_R z}.
\end{align}
Since the contour of the above can be freely deformed, this results in the deformation of the branch cut of the correlator in the complex $\omega$ plane, which is given by solving for the zero of the denominator of the integrand, see \Cref{app:SingularIntegrals} (and also Appendix~A of \cite{Bajec:2024jez}). 

At finite mass $m>0$ the situation changes substantially, since we now have an integral schematically of the form
\begin{align}
    \int_0^1 du\int_{-1}^1dz\,e^{-\frac{m}{T_0}\frac{1}{\sqrt{1-u^2}}}\frac{1}{1-i\omega\tau_R + i k\tau_R \,z u}.
\end{align}
For simplicity, we take $u\in[0,1]$ and deform the angular integration contour $\int_{-1}^1 dz\mapsto\int_{\gam_c} dz$ with a general (smooth) path $\gam_c:[0,1]\rightarrow\mathbb{C}$. We see that the integral is not well-defined for an entire two dimensional area in the complex $\omega$ plane. Indeed, not only must $\omega$ not lie on the contour $\gam_c$, it must also not lie within the region swept by the ray from the origin to the contour, courtesy of $u\in[0,1]$. Deforming the integration contour therefore cuts out a whole region of the complex $\omega$ plane where the correlator is not defined. Different choices of integration contour cut out different regions of the complex plane, which can be made arbitrarily large (see also \cite{Brants:2024wrx} for the discussion on two dimensional areas of non-analyticity in the complex $\omega$ plane). This may be inconsistent with the general properties of the causal correlators.

However, keeping both integration variables $z$ and $u$ on the real axis bypasses the aforementioned problems, since this choice cuts out a finite \textit{curve} in the complex plane, where the correlator is not defined, namely
\begin{align}
    \omega = k\,zu - i/\tau_R,
\end{align}
where $zu \in[-1,1]$. The fact that the angular and momentum/energy integration are non-trivially coupled in the $m>0$ case implies a unique choice of the cut -- the straight line between $\pm k -i/\tau_R$. This is in partial agreement with \cite{Lin:2025ehr}, however we disagree with their interpretation of a `branch point condensate' forming a branch cut. 
In our view, the unique (or preferred) choice of the cut is inconsistent with the  characteristic property of a branch cut, namely the ability to deform its shape. We also note that this non-analytic structure stems solely from the numerator, as the denominator is trivial and analytic in the case of $G^{2,2}_{JJ}$. The change in the analytic structure when $m>0$ is, at least in this case, indeed abrupt and not regulated by the numerator/denominator cancelling effects.

This unique choice for the cut at $m>0$ has deep consequences that were first noticed in the context of massless particles \cite{Romatschke:2015gic}, where it was proposed, that the hydrodynamic (charge diffusion) pole ceases to exist at certain values of $k\tau_R$ as it merges with the branch cut. However, as we discussed above, in the massless case the cut can be freely deformed, in particular in such way, that the pole never reaches the branch cut. On the other hand, in the massive case, the straight cut provides a cutoff on $k$, since the (diffusive) poles of the retarded correlators can reach it at a finite value of the (modulus of the) wavevector, $k_*$, which is analogous to the statement of \cite{Romatschke:2015gic}. Conservatively, the behavior of the pole and cut interacting in the massive case is not clear as it lays outside the domain of the usual theory  and crashing into the cut. We find that the cutoff wavelength $k_*$, at which the poles merge with the cut, i.e. when $\Im \omega = -i/\tau_R$, increases both with mass $m$ and with the number of spatial dimensions $d\geq 2$, although the large mass asymptotics seem to agree for arbitrary $d$.

We interpret this marked contrast between $m=0$ and $m>0$ as a consequence of conformal symmetry breaking in the equation of state.\footnote{We thank Sa{\v s}o Grozdanov for this comment.} It seems natural for a conformal theory ($m=0$) to be in principle valid for arbitrary large $k$. 
The branch cut can in this case be deformed freely, allowing the poles of the correlators to be valid for arbitrary large $k$, as they do not hit the cut for any finite $k$. On the other hand, any finite $m>0$ breaks the conformal symmetry and introduces a cut-off in our theory. This is manifested in the inability to deform the cut. We note again that this inability itself stems from the analytic properties of the retarded correlators.

\section{Conclusion}

In this work, we computed retarded correlators at finite temperature and chemical potential in the case of a massive gas in the RTA. We demonstrated via analytic expressions that the transport coefficients are monotonically decreasing positive functions of mass, which corrects the results of \cite{Hataei:2025mqf}, where they predicted e.g.~the shear diffusion mode becoming negative for some value of mass. Furthermore, as we had the complete expression for the retarded correlators, we were able to extract analytically the profile of the discontinuity of the cut, which we showed corresponded to physical information from the cut, leading to an interpretation of the profile of the discontinuity as  additional causal structure of the correlator. 

The overall picture in the massive case is: the hydrodynamic poles correspond to collective phenomena. The non-hydrodynamic behavior due to the ballistic motion of the particles is captured in the correlator by the cut, which spans between the end points situated at $\omega=\pm k -i/\tau_R$ that indicate the maximal speed of causal propagation. The cut has a particular profile due to the mass, which we show has most of its support between $\omega=\pm v_{\rm cut}k-i/\tau_R$. From a technical perspective, an important take home message is that, as we demonstrated in \Cref{sec:thecut}, multivalued functions stemming from an integral equation should be studied on the level of the integrand in order to avoid confusion about the analytic structure of the result. 

It will be interesting to explore what hidden structures can be uncovered from the  perspective developed here. It seems clear that a natural first step is to consider RTA with an energy dependent relaxation time \cite{Kurkela:2017xis,Kurkela:2018qeb,Kurkela:2019kip,Brants:2024wrx,Hu:2024tnn}. The integration over momenta in this case is more nuanced, since the branch cut of the integrand is being driven deeper in the complex plane for typically studied relaxation times of the form $\tau_R\propto (p^0)^\alpha.$ Moreover, the causal structure of the correlators are largely unexplored in extensions to higher levels in the BBGKY hierarchy \cite{Grozdanov:2024fxr} and in the kinetic theory of  scalar field theory, including in the expanding FLRW background \cite{PhysRevLett.116.022301}. It would be curious to further explore the complex structure in models where it is explicitly known that interactions lead to dynamical lightcones, e.g.~in two fluid model with dynamical effective metrics \cite{Kurkela:2018dku}.

\begin{acknowledgments}
We would like to thank Eduardo Grossi, Sa{\v s}o Grozdanov and Ayan Mukhopadhyay for helpful discussions.
A.S. would like to thank the Galileo Galilei Institute for Theoretical Physics for the hospitality and the INFN for partial support during the completion of this work. A.S. was supported by funding from Horizon Europe research and innovation programme under the Marie Skłodowska-Curie grant agreement No. 101103006 and the project N1-0245 of Slovenian Research Agency (ARIS). M.B. is supported by the research programme P1-0402 of Slovenian Research Agency (ARIS).
\end{acknowledgments}

\appendix

\section{Massive transport in arbitrary dimensions}\label{app:diffusion-arb}

In this appendix, we collect results for transport in dimensions other than 3 spatial dimensions.

\subsection{Charge diffusion}
We first consider the charge diffusion in arbitrary dimensions by focusing on the diffusive pole given by Eq.~\eqref{eq:DiffusivePoleEq}. The integral $I^{00}$ can be generalized to arbitrary dimension:
\begin{align}
    I^{00}_{(d)} = \int \frac{d\Omega^{(d)}}{\Omega^{(d)}} \frac{1}{1 + \tau_R\left( -i \omega + i \gam \textbf{k}\cdot \textbf{v} \right)},
\end{align}
where $d\Omega^{(d)}$ is the $d$--dimensional solid angle differential and $\Omega^{(d)} =\frac{2\pi^{d/2}}{\Gamma(d/2)}$. We define the hyperspherical coordinates in $d$ dimensions in standard fashion:
\begin{align}
    &x^1 = r \cos\theta_1, \\
    &x^2 = r \sin\theta_1 \cos \theta_2, \\
    &\;\;\quad\vdots \nonumber\\
    &x^d = r \sin\theta_1 \ldots \sin \theta_{d-2} \sin \theta_{d-1},
\end{align}
where $\theta_{d-1} \in [0,2\pi)$ and the rest $\theta_i \in [0,\pi)$. The $z$ coordinate in the case $d=3$ would correspond to $x^1$.  The measure is given by
\begin{align}
    d\Omega^{(d)} = \sin^{d-2}\theta_1 \sin^{d-3}\theta_2 \ldots \sin^2\theta_{d-3}\sin\theta_{d-2} d\theta_1\ldots d\theta_{d-1}.
\end{align}
Let us choose our wave vector to be aligned with the $x^1$ axis, i.e. $\textbf{k}\cdot \textbf{v} = k \cos\theta_1$. Then
\begin{align}
    I_{(d)}^{00} &= \frac{\Omega^{(d-1)}}{\Omega^{(d)}} \int_0^\pi d\theta \frac{\sin^{d-2}\theta }{1 - i \omega \tau_R + i \gam k \tau_R \cos\theta} \\
    &=\frac{\Omega^{(d-1)}}{\Omega^{(d)}} \frac{i 2^{d-2} \Gamma \left(\frac{d-1}{2}\right)^2 \,
   _2\tilde{F}_1\left(1,\frac{d-1}{2};d-1;\frac{2 k \gam 
   \tau_R}{k \gam  \tau_R+\omega 
   \tau_R+i}\right)}{\gam  k \tau_R+\tau_R \omega +i},
\end{align}
where $_2\tilde{F}_1$ denotes the hypergeometric regularized function  defined by,
\begin{align}
    _p\tilde{F}_q\equiv\frac{_p F_q\left(a_1,\ldots,a_p;b_1\ldots, b_q;x\right)}{\Gamma(b_1)\ldots\Gamma(b_q)}
\end{align}
and $_p F_q$ is the hypergeometric function \cite{Abramowitz1964}. Then the charge diffusion equation \eqref{eq:DiffusivePoleEq} becomes
\begin{align}
    \int_0^\infty dp\, p^{d-1} \frac{f_0}{T_0} \left(1-I^{00}_{(d)}\right) = 0.
\end{align}
As explained in the main text, we expand the above equation in the hydrodynamic limit to extract the dispersion relation $\omega = - i D_{(d)}(m)k^2+\mathcal{O}(k^4)$:
\begin{align}\label{eq:ChargeDiff_dDim}
    \int_m^\infty dx \frac{e^{-\frac{x}{T_0}}}{x}
   \left(x^2-m^2\right)^{\frac{d-2}{2}} \left(
   D_{(d)} x^2-\frac{\tau_R}{d} \left( x^2-m^2\right)\right)=0.
\end{align}
Clearly the above equation implies $D_{(d)}(m=0) = \tau_R/d$ as expected as well as $D_{(d)}(m)\geq 0$ for any $m\geq0$ and any $d\geq2$. We explicitly write out the result for $d=2$:
\begin{align}
    \frac{D_{(2)}}{\tau_R} = \frac{1}{2}+ \m^2\frac{  e^{m/T_0}
   \text{Ei}\left(-\m\right)}{2
    \left(1+\m\right)},
\end{align}
where $\mathrm{Ei}(z)$ denotes the exponential integral function
\begin{align}\label{eq:eiz}
    \mathrm{Ei}(-z) \equiv - \int_z^\infty dt \frac{e^{-t}}{t}.
\end{align}
The difference between the $d=3$ case in the main text is most noticeable in the asymptotic limit, where the $d=2$ case obtains logarithmic corrections to the massless result:
\begin{align}
    \frac{D_{(2)}}{\tau_R} = \begin{cases}
        \frac{1}{2} +\frac{\m^2}{2} \left(\log \left(\m^2\right)+\gam_\mathrm{E} \right) + \mathcal{O}\left(\m^3\right),& \m\ll 1,\\
   \frac{1}{\m} - 2 \frac{1}{\m^2} + \mathcal{O}\left( \m^{-3}\right),& \m \gg 1,
    \end{cases}
\end{align}
where $\gam_\mathrm{E}$ denotes the Euler-Mascheroni constant.

Additionally, we find in $d=2$ another gapped mode given by
\begin{align}
    \omega = -\frac{2i}{\tau_R} + i D_{(2)} k^2 + \mathcal{O}\left(k^3\right).
\end{align}
The above dispersion relation is analogous to the massless case \cite{Bajec:2024jez}. In this case however, the two poles do not seem to collide, as the indeformable cut stands between them. We comment on this in \Cref{sec:deform}.

\subsection{Shear diffusion}

Start with the solution of the linearized Boltzmann equation in the case of external metric perturbation:
\begin{align}
    \delta f &= \frac{f_0}{T_0} \frac{p^i \delta u_i + \frac{\delta T}{T_0} p^0  - \tau_R \Gamma^0_{\ab}p^0 \tilde{v}^\alpha \tilde{v}^\beta }{1 + \tau_R \left( -i \omega + i \gam \textbf{k}\cdot \textbf{v}  \right)}\\
    &= \frac{f_0}{T_0} \frac{p^\perp \delta u_\perp - \tau_R \Gamma^0_\ab \tilde{v}^\alpha \tilde{v}^\beta}{1 + \tau_R(-i\omega + i k \gam \cos\theta_1)},
\end{align}
where we turned off the perturbations that decouple from the shear channel in the last equality and denoted by $\perp$ any coordinate direction perpendicular to the wave vector which we take to be along the $x^1$ direction. For the sake of concreteness let us pick $p^\perp$ to be aligned along the $x^2$ axis and denote it $p^y$ in order to not confuse it with the square of the momentum. The diffusive pole in the shear channel is  a consequence of the matching condition $\delta T^{0y} = (\varepsilon_0+P_0) \delta u^y$ and is therefore encoded in the following integral equation:
\begin{align}
    1 - \frac{1}{\varepsilon_0+P_0}\int_0^\infty \frac{dp}{(2\pi)^d} \Omega^{(d)} p^{d-1}\left(p^0\right)^2  \frac{f_0}{T_0} I^{22}_{(d)}(p)=0,
\end{align}
where $I^{22}_{(d)}$ is the $d$--dimensional generalization of $I^{22}$ in \cref{eq:iab}:
\begin{align}
    I^{22}_{(d)}&\equiv\int\frac{d\Omega^{(d)}}{\Omega^{(d)}} \frac{\left(\tilde{v}^y\right)^2}{1 + \tau_R\left(-i\omega + i k\gam\cos\theta_1\right)}\\
    &=\gam^2 \frac{i \Gamma (d) \,
   _2\tilde{F}_1\left(1,\frac{d+1}{2};d+1;\frac{2 k
   \gam  \tau_R}{k \gam  \tau_R+\omega  \tau_R+i}\right)}{\gam  k
   \tau_R+\tau_R \omega +i}.
\end{align}
Using $\varepsilon_0 +P_0 = \frac{\partial T^{0y}}{\partial u^y}$, we obtain\footnote{$\int\frac{d\Omega^{(d)}}{\Omega^{(d)}} (v^y)^2 = 1/d$.}
\begin{align}
    \int_0^\infty dp \,p^{d-1} \left(p^0\right)^2 \frac{f_0}{T_0}\left(\frac{\gam^2}{d} - I^{22}_{(d)}\right) = 0.
\end{align}
Expanding again in the hydrodynamic limit, we can obtain the equation for the shear diffusion constant $\omega = -i D^{(d)}_\mathrm{sh}(m) k^2 + \mathcal{O}(k^4)$:
\begin{align}
    \int_m^\infty dx \frac{ e^{-\frac{x}{T_0}}}{x}
   \left(x^2-m^2\right)^{d/2} \left( D_\mathrm{sh}^{(d)}
   x^2-\frac{\tau_R}{d+2} (x^2-m^2)\right) = 0.
\end{align}
The limit $m\rightarrow 0$ implies $D^{(d)}_\mathrm{sh}/\tau_R = 1/(d+2)$, as well as $D^{(d)}_\mathrm{sh}\geq0$ for any mass $m\geq0$ and any $d\geq2$. We note that this result also includes a general result for the shear viscosity through the hydrodynamic relation $D_\mathrm{sh} = \eta / (\varepsilon_0 +P_0)$.

We write out the $d=2$ case explicitly
\begin{align}
    D_\mathrm{sh}^{(2)}/\tau_R=\frac{ \m^4\,e^{\m}
   \Gamma
   (0,\m)+\left(-\m^2+\m+6\right
   ) \m+6}{8 (\m
   (\m+3)+3)},
\end{align}
where $\Gamma(0,z)$ is the incomplete gamma function and $\m=m/T_0$. The asymptotics are given by
\begin{align}
    D^{(2)}_\mathrm{sh}(m)/\tau_R = \begin{cases}
        \frac{1}{4}-\frac{1}{24}\m^2+ \mathcal{O}\left(\m^3\right), & \m\ll 1,\\
   \frac{1}{\m} - 3 \frac{1}{\m^2} + \mathcal{O}\left( \m^{-3}\right),& \m \gg 1.
    \end{cases}
\end{align}

\section{Discontinuity profile, branch cuts and branch points}\label{app:discontinuity}

\subsection{Integrals of functions with a branch cut}\label{app:DiscontinuityProfile}

In this appendix we showcase the general procedure to describe the analytic structure of an integral of a function with branch points via several examples of increasing complexity.

\begin{figure}[tbp]
    \centering
    \includegraphics[width=0.49\linewidth]{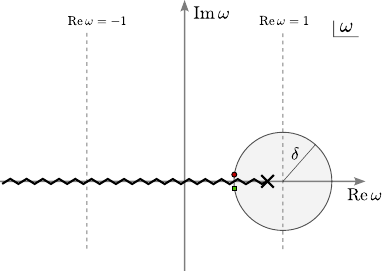}
    \caption{The analytic structure of the integrand of \eqref{eq:Example1} at a fixed $z_0 \in [-1,-1+\delta)$. The position of $\omega^\pm$ is represented by a red circle and green square respectively. As mentioned in the main text, the difference $f(\omega_+;z_0)-f(\omega_-;z_0)$, in the limit when the two points approach each other, is $2\pi i$ when the branch cut is separating them and zero otherwise.}
    \label{fig:discontinuity11}
\end{figure}

\textit{Example 1. } We start with the simplest example\footnote{Note that we keep the definition of $F(\omega)$ local to each example to avoid introducing excess notation.}
\begin{align}\label{eq:Example1}
    F(\omega)\equiv \int_{-1}^1dz\log(z+\omega) \equiv \int_{-1}^1 dz\, f(z;\omega),
\end{align}
where we choose the branch cut of the complex logarithm $\log z$ to lie along the negative real axis $(-\infty,0]$. We take the domain of $\omega \in \Omega\subseteq\mathbb{C}$ to be such that the above integral is well defined. We will later analytically continue $\omega$ to the whole complex plane. As $z$ varies from $-1$ to $1$, the branch point of the integrand $\log(z+\omega)$ varies from $\omega = 1$ to $\omega = -1$. In order to determine the discontinuity profile at $\omega_*$, we define $\omega_\pm\equiv \omega_* \pm i \varepsilon$ for some $\varepsilon>0$. The purely imaginary displacement is due to the choice of the branch cut, which is aligned with the real axis. Then, for $z=z_0$:
\begin{align}
    \mathrm{disc}\, f(z_0;\omega_*) \equiv \lim_{\varepsilon\rightarrow 0}\left( f(z_0;\omega_+) - f(z_0;\omega_-) \right) 
    &=\begin{cases}
        2\pi i, \quad \omega_\pm\text{ are separated by a branch cut},\\
        0,\quad\text{otherwise}.
    \end{cases}
\end{align}
The integral over $z$ of the above is the discontinuity profile of the integrated function $F(\omega)$. In practice this reduces to integrating $2\pi i$ over the interval of $z$, for which the branch cut separates $\omega_\pm$. We position ourselves at the furthermost branch point of the integrand $f(z,\omega)$, denoted $\omega=\omega_\mathrm{bp}$, which we define to be the branch point which shares the branch cut with all the other branch points as we vary $z$ over the integration interval. In the present case, this means $\omega_\mathrm{bp} = 1$, when $z_0 = -1$. The discontinuity profile at $\omega = \omega_\mathrm{bp}-\delta=1 -\delta$ is then the integral of $2\pi i$, over such interval of $z$, that the distance of the branch point of the integrand $f(z,\omega)$ from the furthermost branch point is not more than $\delta$ (see Fig.~\ref{fig:discontinuity11}).
We note that this description with $\delta$ is a choice of personal convenience. Therefore, the discontinuity of $F(\omega)$ is given by
\begin{align}
    \mathrm{disc}\,F(\omega)\Big\vert_{\omega \in[-1,1]} = 2\pi i \int_{-1}^{-1+\delta}dz = 2\pi i \,\delta
\end{align}
This result is valid for $\delta \in [0,2]$. For $\delta>2$, the discontinuity is constant $2\pi i\int_{-1}^{1}dz = 4\pi i$. Therefore, the final analytic structure of $F(\omega)$ is a branch cut on the real axis spanning $(-\infty,1]$, with a linear discontinuity profile in $[-1,1]$. This is in contrast with, for instance, the discontinuity of the complex logarithm, where the discontinuity is constant along the whole branch cut.

We mention that this analytic structure is compatible with the `naive' integration
\begin{align}
    F(\omega)&=(1+\omega)\log(1+\omega)-(1+\omega)-(-1+\omega)\log(-1+\omega)+(-1+\omega),\\
    &=\omega\log\frac{\omega+1}{\omega-1}+\log(\omega+1)+\log(\omega-1) +2,
\end{align}
where the first term has a linear profile in the discontinuity 
between $-1$ and $1$ (due to the factor of $\omega$),  
while the sum of the following two terms lead to the the discontinuity of $4\pi i$ for $\omega < -1$.

\begin{figure}
    \centering
   \includegraphics[width=0.49\linewidth]{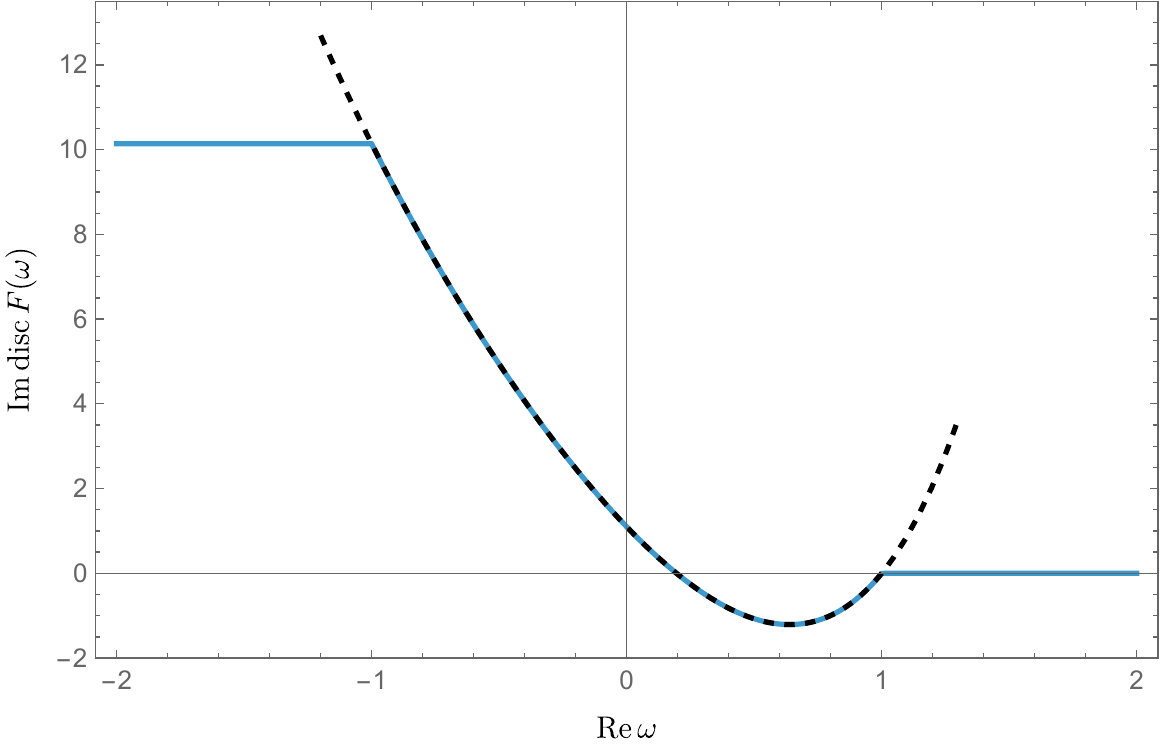}
    \caption{An example of discontinuity profile given by Eq.~\eqref{eq:DiscontinuityFormula2} for $f(z,\omega) = z^2 + e^{-z}z+\cos z$. The numerical result is shown in blue, which is obtained by numerically integrating \eqref{eq:Example2Integral}. The analytic prediction, given by Eq.~\eqref{eq:Example2_Analytic2}, is represented by the dashed black line.
    \label{fig:discontinuityExample22}}
\end{figure}

\textit{Example 2.} The above example naturally generalizes to include functions of the form
\begin{align}\label{eq:Example2Integral}
    F(\omega) = \int_{-1}^1dz f(z,\omega) \log(z+\omega),
\end{align}
where $f(z,\omega)$ is any (holomorphic) function. Then the same argument as above implies the following discontinuity profile between $-1$ and $1$:
\begin{align}\label{eq:DiscontinuityFormula2}
    \text{disc }F(\omega)\Big\vert_{\omega= 1-\delta}= 2\pi i \int_{-1}^{-1+\delta}dz f(z,\omega)\big\vert_{\omega= 1-\delta}.
\end{align}
Taking a concrete example of
\begin{align}
    f(z,\omega) = z^2 + e^{-z}z+\cos z,
\end{align}
evaluating the above integral and substituting $\delta = 1-\omega$, we obtain the following result
\begin{align}\label{eq:Example2_Analytic2}
    \mathrm{disc}\,F(\omega) = \frac{2}{3} i \pi  \left(1+3 \sin (1)-\omega ^3+3 e^{\omega } (\omega -1)-3 \sin (\omega )\right),
\end{align}
where $\omega \in[-1,1]\subseteq\mathbb{R}$. We provide the numerical check of the above result in Fig.~\ref{fig:discontinuityExample22}.

\textit{Example 3.} We conclude with the final example, closely related to the integral found in our kinetic theory set up. Consider now\footnote{One can reinstate a finite $\tau_R$ by making a substitution $\omega \mapsto\omega + i/\tau_R$.}
\begin{align}
    F(\omega;k,m) = \int_m^\infty dz\, e^{-\frac{z}{T_0}} \log\left(\omega -k\sqrt{1-\frac{m^2}{z^2}}\right).
\end{align}
Here we also explicitly state the method to extract the discontinuity profile:
\begin{enumerate}
    \item Choose a branch cut of the integrand and consider all of its branch points (and cuts) as the integration parameter varies. Identify the `furthermost' branch point $\omega_\mathrm{bp}$, meaning the branch point that shares the branch cut with all the other branch points (when varying the integration parameter).

    For the present case, we take $\log z$ to have a cut along the real axis $(-\infty,0]$. Therefore the furthermost branch point is $\omega_\mathrm{bp} = k$, when $z\rightarrow \infty$. We make the substitution $z = \frac{m}{u}$ in order to bring infinity to a finite value:
    \begin{align}
        F(\omega;m,k) = m \int_0^1 \frac{du}{u^2} e^{-\frac{m}{T_0}\frac{1}{u}} \log\left( \omega - k\sqrt{1-u^2} \right).
    \end{align}
    Now the furthermost branch point corresponds to the value of $u=0$.
    \item Fix a value $\delta>0$ and calculate the interval of $u$ for which the branch point of the integrand is still less than $\delta$ away from $\omega_\mathrm{bp}$. In our case, the limiting value of $u$ is given by
    \begin{align}
        -k\sqrt{1-u_\delta^2}=\varepsilon\Longrightarrow u_\delta = \sqrt{1-\frac{\delta^2}{k^2}}.
    \end{align}
    \item Finally integrate the discontinuity (in this case $2\pi i$ because of the logarithm) along the calculated interval:
    \begin{align}     \mathrm{disc}\,F(\omega;k,m)\Big\vert_{\omega=\omega_{\mathrm{bp}}-\delta}&=2\pi i\, m\int_0^{u_\delta = \sqrt{1-\frac{\delta^2}{k^2}}} \frac{du}{u^2} e^{-\frac{m}{T_0}\frac{1}{u}} \\
       &=2  \pi i \,T_0\exp\left( -\frac{m}{T_0} \frac{1}{\sqrt{1-\frac{\delta^2}{k^2}}} \right).\label{eq:FinalAnalytic}
    \end{align}
    The final step is a trivial transformation from $\delta$ to $\omega$.
\end{enumerate}
We can use the analytic profile \cref{eq:FinalAnalytic} to establish a criterion for an effective velocity. After normalizing, we want to define a scale, $0<\epsilon<1$, which distinguishes between the dominant and the exponentially suppressed part of the branch cut. Namely, 
\begin{align}\label{eq:cutoff}
    e^{\frac{m}{T_0}\left(1-(1-\frac{\omega^2}{k^2})^{-1/2}\right)}=\epsilon.
\end{align}
Solving for $\omega$, we find
\begin{align}\label{eq:vcut}
    \omega = \pm\frac{ \sqrt{ (\log \epsilon -2 m)\log \epsilon}}{m-\log \epsilon }k\equiv \pm v_{\rm cut} k.
\end{align}
Note that the $\epsilon\rightarrow0$ limit leads to $v_{\rm cut}=1.$

\begin{figure}
    \centering
    \includegraphics[width=0.55\linewidth]{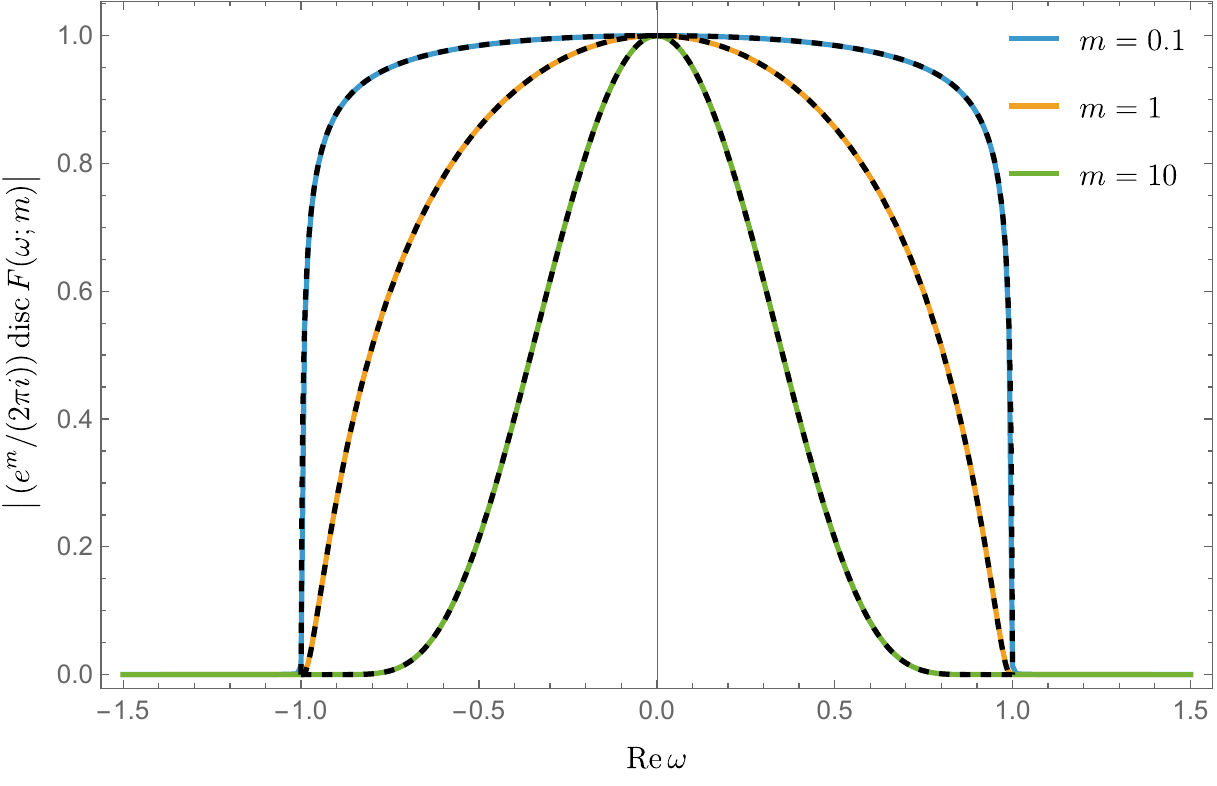}
    \caption{The (normalized) discontinuity profile of Eq.~\eqref{eq:FinalExample} for different values of mass $m$. We set $T_0 k=1$. The analytic prediction of \eqref{eq:FinalAnalytic} is represented by the dashed black lines.}
    \label{fig:massiveCutExample}
\end{figure}

We note here that the final result is symmetric in $\omega$ and can be applied also in the case of 
\begin{align}\label{eq:FinalExample}
    F(\omega;k,m) = \int_m^\infty dz\, e^{-\frac{z}{T_0}} \log\left(\frac{\omega -k\sqrt{1-\frac{m^2}{z^2}}}{\omega +k\sqrt{1-\frac{m^2}{z^2}}}\right).
\end{align}
We confirm numerically that the previous expression is captured by the analytic expression \cref{eq:FinalAnalytic} in Fig.~\ref{fig:massiveCutExample}.

\begin{figure}
    \centering
    \includegraphics[width=0.95\linewidth]{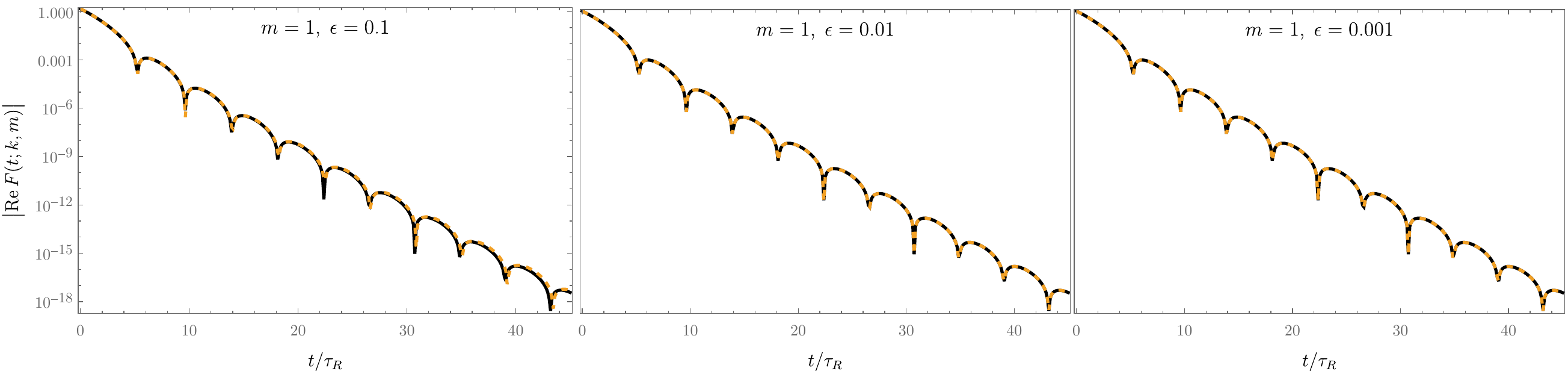}
    \includegraphics[width=0.95\linewidth]{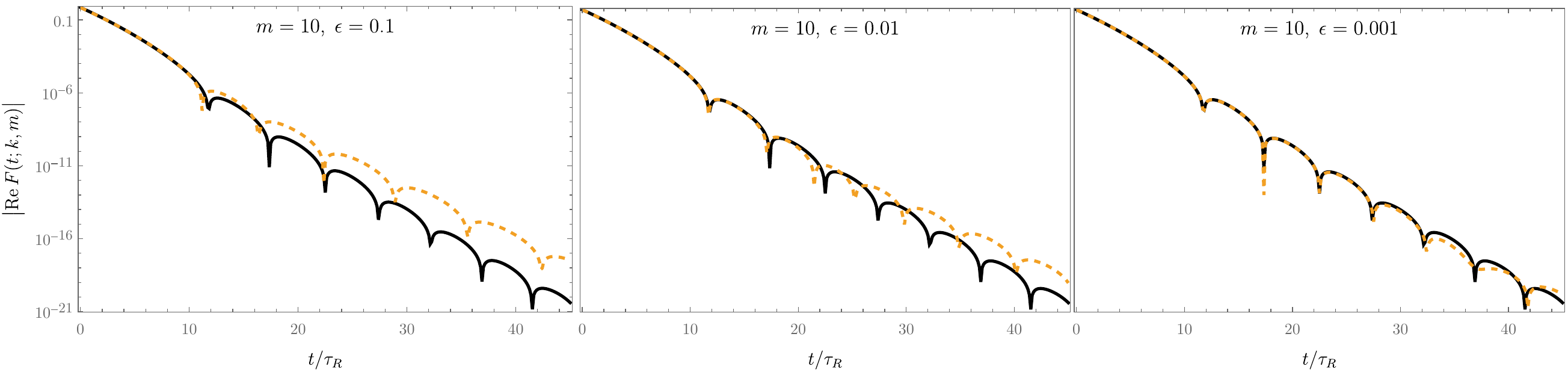}
    \caption{Absolute value of the real part of the  Fourier transform of \cref{eq:FinalExample}. The solid color denotes the exact result, i.e. taking the entire branch cut, while the dashed line is only integrating the part of the branch cut with the most support, parameterized by  $\epsilon$ in \cref{eq:vcut}.}
    \label{fig:fourier-transform}
\end{figure}

Finally, similar to the discussion in \Cref{sec:thecut}, we will discuss the Fourier transform of \cref{eq:FinalExample} by integrating in two different ways as shown in \cref{fig:dogbone}. We show the result of these different procedures in \Cref{fig:fourier-transform} via the absolute value of the real part of the Fourier transform of \cref{eq:FinalExample} for two different masses $m/T_0=1,15$ and for a variety of cutoffs, defined in \cref{eq:cutoff}. We see that there is excellent agreement for $\epsilon=10^{-3}$ over multiple oscillations and relaxation times, which becomes worse for increasing mass and at late times.

\subsection{Complex integrals of meromorphic functions}\label{app:SingularIntegrals}

Here we review and expand upon the discussion from Appendix A found in \cite{Bajec:2024jez}, which supports the discussion in \Cref{sec:deform}. For the sake of completeness we restate the origin of the branch cut structure in RTA kinetic theory. We proceed in similar fashion to Appendix \ref{app:DiscontinuityProfile} by first considering simple examples.

\begin{figure}[tbp]
    \centering
    \includegraphics[width=0.49\linewidth]{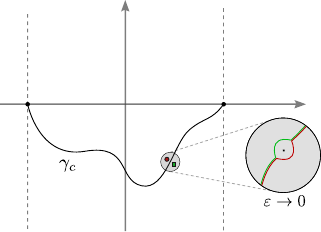}
    \caption{A complex integration contour $\gam_c$ with $\omega_\pm$ schematically represented by the red circle and green square, respectively. An equivalent picture, in the limit where we send the two points to the integration contour, is a simple infinitesimal deformation of the integration contour either above (green contour) or below (red contour) the chosen point on the contour $\omega_*$ (in the plot denoted as a black dot between the two contours in color). The difference between the two choices is then, by the residue theorem, $2\pi i$ times the residue of the integrand at $\omega_*$.}
    \label{fig:singular_contour_int}
\end{figure}

\textit{Example 1.} The prototypical example of an integral sharing key features with computations encountered in the physical of kinetic theory in RTA approximation is
\begin{align}\label{eq:SingInt1}
    F(\omega) \equiv\int_{-1}^{1}dz \frac{1}{z-\omega}.
\end{align}
Treating the integration variable as complex, we may deform the above integration contour arbitrarily into the complex plane. The above integral is then defined for any $\omega$, which does not lie on the integration contour, as per the usual complex analysis arguments, cf.~\cite{Ahlfors1966}. This results in the domain of $F:\Omega\rightarrow\mathbb{C}$ to be a cut complex plane, with the cut given by the integration contour $\gam_c$, i.e. $\Omega = \mathbb{C}\setminus\mathrm{im}\,\gam_c$. This is the basis of the argument why the logarithmic branch cut can be arbitrarily deformed in the massless case. The discontinuity of the function $F(\omega)$ is deduced from the residue theorem as follows --  we position ourselves with our parameter $\omega$ just above and below a point on the contour $\omega_*$ to define $\omega_\pm = \omega_* \pm i\varepsilon$ for some infinitesimal $\varepsilon>0$.\footnote{For simplicity we assume that the tangent to the integration contour at $\omega_*$ is not parallel to the imaginary axis. Otherwise we infinitesimally displace (also) the real parts.} As we send $\varepsilon\rightarrow 0$, we see that the difference between the two values is precisely $2\pi i$ times the residue of the integrand at the point $\omega_*$ (see \Cref{fig:singular_contour_int}).

In the simple case of \Cref{eq:SingInt1} the discontinuity is constant $2\pi i$ along the contour of integration, which agrees with the result if we integrate \Cref{eq:SingInt1} to obtain
\begin{align}
    F(\omega) = \log\frac{\omega+1}{\omega-1}.
\end{align}

\textit{Example 2.} It is then clear that the generalization of the above example to
\begin{align}
    F(\omega) \equiv \int_{\gam_c} dz\frac{f(z,\omega)}{z-\omega},
\end{align}
with $f(z,\omega)$ holomorphic on the integration contour (or rather in its neighborhood), results in the discontinuity profile of $f(z=\omega_*,\,\omega=\omega_*)$, where $\omega_*$ runs along the integration contour; $\omega_*\in\mathrm{im}\,\gam_c$.

\textit{Example 3.} We conclude with an example applicable to the discussion in the main text, where we discuss the discontinuity profile at the branch cut resulting from energy integration at a fixed angle:
\begin{align}\label{eq:SingIntegralEx3}
    F(\omega;k,m) \equiv \int_m^\infty dx\,e^{-\frac{x}{T_0}} \frac{1}{1 + \tau_R\left(-i\omega + i k \sqrt{1-\frac{m^2}{x^2}}\cos\theta\right)}.
\end{align}
The derivation is clearer if we introduce a new variable $u = \sqrt{1-\frac{m^2}{x^2}}$:
\begin{align}
    F(\omega;k,m) = \frac{1}{i k  \tau_R \cos\theta}\int_0^1 du \frac{m u}{(1-u^2)^{3/2}}\exp\left( -\frac{m}{T_0} \frac{1}{\sqrt{1-u^2}} \right) \frac{1}{u - \frac{i\omega\tau_R - 1}{i k \tau_R \cos\theta}}.
\end{align}
Taking the integration contour along the real axis ($u\in[0,1]$) results in the cut along
\begin{align}
    \frac{i\omega\tau_R - 1}{i k \tau_R \cos\theta} \in [0,1]\Longleftrightarrow \omega =  s  k\cos\theta - \frac{i}{\tau_R},\quad s\in[0,1].
\end{align}
Then the discontinuity at this cut, for fixed $k$ and $\theta$, is given by
\begin{align}\label{eq:SingularDisc3Analytic}
    \mathrm{disc} \,F(\omega;k,m)\bigg\vert_{\omega=k\cos\theta\cdot s - \frac{i}{\tau_R}} =  \frac{2\pi i}{ik\tau_R \cos\theta} \frac{m \,s}{(1-s^2)^{3/2}}\exp\left( -\frac{m}{T_0} \frac{1}{\sqrt{1-s^2}} \right).
\end{align}
\begin{figure}
    \centering
    \includegraphics[width=0.55\linewidth]{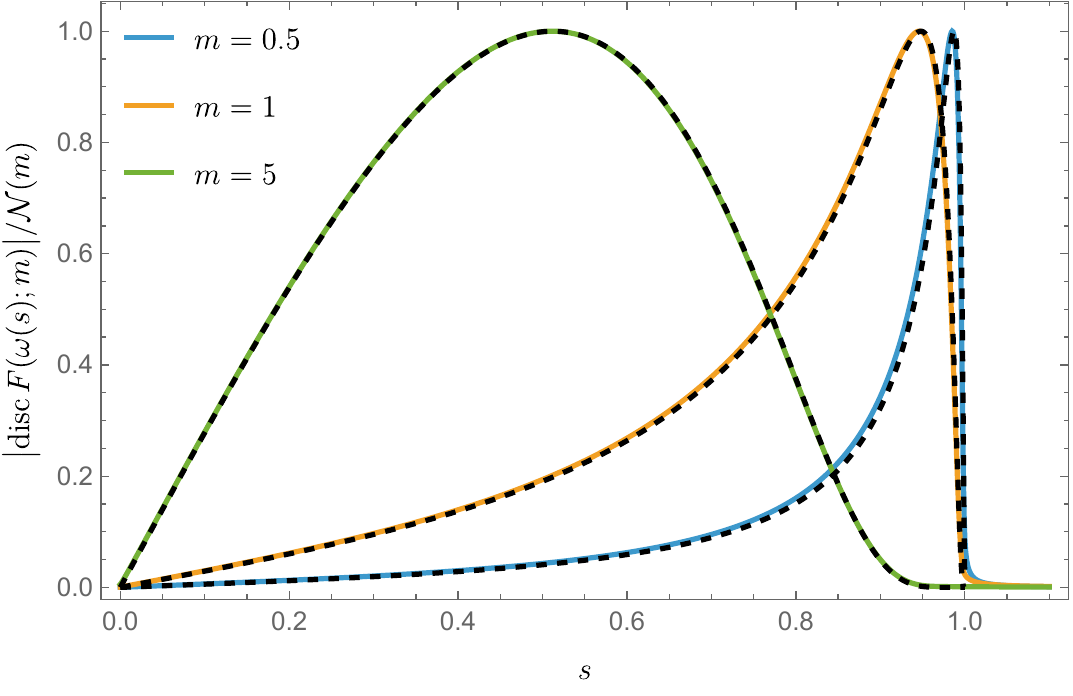}
    \caption{Absolute value of the normalized discontinuity of the integral \eqref{eq:SingIntegralEx3} for different values of mass $m$ as a function of $s=\Re\omega/(k\cos\theta)$ which parametrizes the cut. The black dashed lines denote the analytic result, given by Eq.~\eqref{eq:SingularDisc3Analytic}. We attribute slight misalignment of the numerical and analytic result at small $m$ to numerical error. We set $T_0$, $\tau_R$ and $\cos\theta$ to unity.}
    \label{fig:singularDiscontinuityExample_Check}
\end{figure}
We provide a numerical check of consistency of the above analytic result in Fig.~\ref{fig:singularDiscontinuityExample_Check}.

\bibliography{bib}

\begin{thebibliography}{55}%
\makeatletter
\providecommand \@ifxundefined [1]{%
 \@ifx{#1\undefined}
}%
\providecommand \@ifnum [1]{%
 \ifnum #1\expandafter \@firstoftwo
 \else \expandafter \@secondoftwo
 \fi
}%
\providecommand \@ifx [1]{%
 \ifx #1\expandafter \@firstoftwo
 \else \expandafter \@secondoftwo
 \fi
}%
\providecommand \natexlab [1]{#1}%
\providecommand \enquote  [1]{``#1''}%
\providecommand \bibnamefont  [1]{#1}%
\providecommand \bibfnamefont [1]{#1}%
\providecommand \citenamefont [1]{#1}%
\providecommand \href@noop [0]{\@secondoftwo}%
\providecommand \href [0]{\begingroup \@sanitize@url \@href}%
\providecommand \@href[1]{\@@startlink{#1}\@@href}%
\providecommand \@@href[1]{\endgroup#1\@@endlink}%
\providecommand \@sanitize@url [0]{\catcode `\\12\catcode `\$12\catcode `\&12\catcode `\#12\catcode `\^12\catcode `\_12\catcode `\%12\relax}%
\providecommand \@@startlink[1]{}%
\providecommand \@@endlink[0]{}%
\providecommand \url  [0]{\begingroup\@sanitize@url \@url }%
\providecommand \@url [1]{\endgroup\@href {#1}{\urlprefix }}%
\providecommand \urlprefix  [0]{URL }%
\providecommand \Eprint [0]{\href }%
\providecommand \doibase [0]{http://dx.doi.org/}%
\providecommand \selectlanguage [0]{\@gobble}%
\providecommand \bibinfo  [0]{\@secondoftwo}%
\providecommand \bibfield  [0]{\@secondoftwo}%
\providecommand \translation [1]{[#1]}%
\providecommand \BibitemOpen [0]{}%
\providecommand \bibitemStop [0]{}%
\providecommand \bibitemNoStop [0]{.\EOS\space}%
\providecommand \EOS [0]{\spacefactor3000\relax}%
\providecommand \BibitemShut  [1]{\csname bibitem#1\endcsname}%
\let\auto@bib@innerbib\@empty
\bibitem [{\citenamefont {Pitaevskii}\ and\ \citenamefont {Lifshitz}(2012)}]{pitaevskii2012physical}%
  \BibitemOpen
  \bibfield  {author} {\bibinfo {author} {\bibfnamefont {L.}~\bibnamefont {Pitaevskii}}\ and\ \bibinfo {author} {\bibfnamefont {E.}~\bibnamefont {Lifshitz}},\ }\href@noop {} {\emph {\bibinfo {title} {Physical Kinetics: Volume 10}}},\ \bibinfo {number} {let. 10}\ (\bibinfo  {publisher} {Butterworth-Heinemann},\ \bibinfo {year} {2012})\BibitemShut {NoStop}%
\bibitem [{\citenamefont {de~Groot}\ \emph {et~al.}(1980)\citenamefont {de~Groot}, \citenamefont {van Leeuwen},\ and\ \citenamefont {van Weert}}]{groot-book}%
  \BibitemOpen
  \bibfield  {author} {\bibinfo {author} {\bibfnamefont {S.}~\bibnamefont {de~Groot}}, \bibinfo {author} {\bibfnamefont {W.}~\bibnamefont {van Leeuwen}}, \ and\ \bibinfo {author} {\bibfnamefont {C.}~\bibnamefont {van Weert}},\ }\href@noop {} {\emph {\bibinfo {title} {{Relativistic Kinetic Theory}}}}\ (\bibinfo  {publisher} {North-Holland Publishing Company},\ \bibinfo {address} {Amsterdam, Netherlands},\ \bibinfo {year} {1980})\BibitemShut {NoStop}%
\bibitem [{\citenamefont {Anderson}\ and\ \citenamefont {Witting}(1974)}]{ANDERSON1974466}%
  \BibitemOpen
  \bibfield  {author} {\bibinfo {author} {\bibfnamefont {J.}~\bibnamefont {Anderson}}\ and\ \bibinfo {author} {\bibfnamefont {H.}~\bibnamefont {Witting}},\ }\href {\doibase https://doi.org/10.1016/0031-8914(74)90355-3} {\bibfield  {journal} {\bibinfo  {journal} {Physica}\ }\textbf {\bibinfo {volume} {74}},\ \bibinfo {pages} {466} (\bibinfo {year} {1974})}\BibitemShut {NoStop}%
\bibitem [{\citenamefont {Liboff}(2003)}]{liboff-book}%
  \BibitemOpen
  \bibfield  {author} {\bibinfo {author} {\bibfnamefont {R.}~\bibnamefont {Liboff}},\ }\href@noop {} {\emph {\bibinfo {title} {{Kinetic Theory}}}}\ (\bibinfo  {publisher} {Springer-Verlag},\ \bibinfo {address} {New York, USA},\ \bibinfo {year} {2003})\BibitemShut {NoStop}%
\bibitem [{\citenamefont {Grozdanov}\ \emph {et~al.}(2016)\citenamefont {Grozdanov}, \citenamefont {Kaplis},\ and\ \citenamefont {Starinets}}]{Grozdanov:2016vgg}%
  \BibitemOpen
  \bibfield  {author} {\bibinfo {author} {\bibfnamefont {S.}~\bibnamefont {Grozdanov}}, \bibinfo {author} {\bibfnamefont {N.}~\bibnamefont {Kaplis}}, \ and\ \bibinfo {author} {\bibfnamefont {A.~O.}\ \bibnamefont {Starinets}},\ }\href {\doibase 10.1007/JHEP07(2016)151} {\bibfield  {journal} {\bibinfo  {journal} {JHEP}\ }\textbf {\bibinfo {volume} {07}},\ \bibinfo {pages} {151} (\bibinfo {year} {2016})},\ \Eprint {http://arxiv.org/abs/1605.02173} {arXiv:1605.02173 [hep-th]} \BibitemShut {NoStop}%
\bibitem [{\citenamefont {Romatschke}(2016)}]{Romatschke:2015gic}%
  \BibitemOpen
  \bibfield  {author} {\bibinfo {author} {\bibfnamefont {P.}~\bibnamefont {Romatschke}},\ }\href {\doibase 10.1140/epjc/s10052-016-4169-7} {\bibfield  {journal} {\bibinfo  {journal} {Eur. Phys. J. C}\ }\textbf {\bibinfo {volume} {76}},\ \bibinfo {pages} {352} (\bibinfo {year} {2016})},\ \Eprint {http://arxiv.org/abs/1512.02641} {arXiv:1512.02641 [hep-th]} \BibitemShut {NoStop}%
\bibitem [{\citenamefont {Florkowski}\ \emph {et~al.}(2013{\natexlab{a}})\citenamefont {Florkowski}, \citenamefont {Ryblewski},\ and\ \citenamefont {Strickland}}]{Florkowski:2013lza}%
  \BibitemOpen
  \bibfield  {author} {\bibinfo {author} {\bibfnamefont {W.}~\bibnamefont {Florkowski}}, \bibinfo {author} {\bibfnamefont {R.}~\bibnamefont {Ryblewski}}, \ and\ \bibinfo {author} {\bibfnamefont {M.}~\bibnamefont {Strickland}},\ }\href {\doibase 10.1016/j.nuclphysa.2013.08.004} {\bibfield  {journal} {\bibinfo  {journal} {Nucl. Phys. A}\ }\textbf {\bibinfo {volume} {916}},\ \bibinfo {pages} {249} (\bibinfo {year} {2013}{\natexlab{a}})},\ \Eprint {http://arxiv.org/abs/1304.0665} {arXiv:1304.0665 [nucl-th]} \BibitemShut {NoStop}%
\bibitem [{\citenamefont {Florkowski}\ \emph {et~al.}(2013{\natexlab{b}})\citenamefont {Florkowski}, \citenamefont {Ryblewski},\ and\ \citenamefont {Strickland}}]{Florkowski:2013lya}%
  \BibitemOpen
  \bibfield  {author} {\bibinfo {author} {\bibfnamefont {W.}~\bibnamefont {Florkowski}}, \bibinfo {author} {\bibfnamefont {R.}~\bibnamefont {Ryblewski}}, \ and\ \bibinfo {author} {\bibfnamefont {M.}~\bibnamefont {Strickland}},\ }\href {\doibase 10.1103/PhysRevC.88.024903} {\bibfield  {journal} {\bibinfo  {journal} {Phys. Rev. C}\ }\textbf {\bibinfo {volume} {88}},\ \bibinfo {pages} {024903} (\bibinfo {year} {2013}{\natexlab{b}})},\ \Eprint {http://arxiv.org/abs/1305.7234} {arXiv:1305.7234 [nucl-th]} \BibitemShut {NoStop}%
\bibitem [{\citenamefont {Florkowski}\ \emph {et~al.}(2014)\citenamefont {Florkowski}, \citenamefont {Maksymiuk}, \citenamefont {Ryblewski},\ and\ \citenamefont {Strickland}}]{Florkowski:2014sfa}%
  \BibitemOpen
  \bibfield  {author} {\bibinfo {author} {\bibfnamefont {W.}~\bibnamefont {Florkowski}}, \bibinfo {author} {\bibfnamefont {E.}~\bibnamefont {Maksymiuk}}, \bibinfo {author} {\bibfnamefont {R.}~\bibnamefont {Ryblewski}}, \ and\ \bibinfo {author} {\bibfnamefont {M.}~\bibnamefont {Strickland}},\ }\href {\doibase 10.1103/PhysRevC.89.054908} {\bibfield  {journal} {\bibinfo  {journal} {Phys. Rev. C}\ }\textbf {\bibinfo {volume} {89}},\ \bibinfo {pages} {054908} (\bibinfo {year} {2014})},\ \Eprint {http://arxiv.org/abs/1402.7348} {arXiv:1402.7348 [hep-ph]} \BibitemShut {NoStop}%
\bibitem [{\citenamefont {Kurkela}\ and\ \citenamefont {Wiedemann}(2019)}]{Kurkela:2017xis}%
  \BibitemOpen
  \bibfield  {author} {\bibinfo {author} {\bibfnamefont {A.}~\bibnamefont {Kurkela}}\ and\ \bibinfo {author} {\bibfnamefont {U.~A.}\ \bibnamefont {Wiedemann}},\ }\href {\doibase 10.1140/epjc/s10052-019-7271-9} {\bibfield  {journal} {\bibinfo  {journal} {Eur. Phys. J. C}\ }\textbf {\bibinfo {volume} {79}},\ \bibinfo {pages} {776} (\bibinfo {year} {2019})},\ \Eprint {http://arxiv.org/abs/1712.04376} {arXiv:1712.04376 [hep-ph]} \BibitemShut {NoStop}%
\bibitem [{\citenamefont {Kurkela}\ \emph {et~al.}(2019{\natexlab{a}})\citenamefont {Kurkela}, \citenamefont {Wiedemann},\ and\ \citenamefont {Wu}}]{Kurkela:2019kip}%
  \BibitemOpen
  \bibfield  {author} {\bibinfo {author} {\bibfnamefont {A.}~\bibnamefont {Kurkela}}, \bibinfo {author} {\bibfnamefont {U.~A.}\ \bibnamefont {Wiedemann}}, \ and\ \bibinfo {author} {\bibfnamefont {B.}~\bibnamefont {Wu}},\ }\href {\doibase 10.1140/epjc/s10052-019-7428-6} {\bibfield  {journal} {\bibinfo  {journal} {Eur. Phys. J. C}\ }\textbf {\bibinfo {volume} {79}},\ \bibinfo {pages} {965} (\bibinfo {year} {2019}{\natexlab{a}})},\ \Eprint {http://arxiv.org/abs/1905.05139} {arXiv:1905.05139 [hep-ph]} \BibitemShut {NoStop}%
\bibitem [{\citenamefont {Dash}\ \emph {et~al.}(2022)\citenamefont {Dash}, \citenamefont {Bhadury}, \citenamefont {Jaiswal},\ and\ \citenamefont {Jaiswal}}]{Dash:2021ibx}%
  \BibitemOpen
  \bibfield  {author} {\bibinfo {author} {\bibfnamefont {D.}~\bibnamefont {Dash}}, \bibinfo {author} {\bibfnamefont {S.}~\bibnamefont {Bhadury}}, \bibinfo {author} {\bibfnamefont {S.}~\bibnamefont {Jaiswal}}, \ and\ \bibinfo {author} {\bibfnamefont {A.}~\bibnamefont {Jaiswal}},\ }\href {\doibase 10.1016/j.physletb.2022.137202} {\bibfield  {journal} {\bibinfo  {journal} {Phys. Lett. B}\ }\textbf {\bibinfo {volume} {831}},\ \bibinfo {pages} {137202} (\bibinfo {year} {2022})},\ \Eprint {http://arxiv.org/abs/2112.14581} {arXiv:2112.14581 [nucl-th]} \BibitemShut {NoStop}%
\bibitem [{\citenamefont {Ambrus}\ \emph {et~al.}(2023)\citenamefont {Ambrus}, \citenamefont {Schlichting},\ and\ \citenamefont {Werthmann}}]{Ambrus:2022koq}%
  \BibitemOpen
  \bibfield  {author} {\bibinfo {author} {\bibfnamefont {V.~E.}\ \bibnamefont {Ambrus}}, \bibinfo {author} {\bibfnamefont {S.}~\bibnamefont {Schlichting}}, \ and\ \bibinfo {author} {\bibfnamefont {C.}~\bibnamefont {Werthmann}},\ }\href {\doibase 10.1103/PhysRevD.107.094013} {\bibfield  {journal} {\bibinfo  {journal} {Phys. Rev. D}\ }\textbf {\bibinfo {volume} {107}},\ \bibinfo {pages} {094013} (\bibinfo {year} {2023})},\ \Eprint {http://arxiv.org/abs/2211.14379} {arXiv:2211.14379 [hep-ph]} \BibitemShut {NoStop}%
\bibitem [{\citenamefont {Brants}(2024)}]{Brants:2024wrx}%
  \BibitemOpen
  \bibfield  {author} {\bibinfo {author} {\bibfnamefont {R.}~\bibnamefont {Brants}},\ }\href {\doibase 10.1103/PhysRevD.110.116027} {\bibfield  {journal} {\bibinfo  {journal} {Phys. Rev. D}\ }\textbf {\bibinfo {volume} {110}},\ \bibinfo {pages} {116027} (\bibinfo {year} {2024})},\ \Eprint {http://arxiv.org/abs/2409.09022} {arXiv:2409.09022 [hep-th]} \BibitemShut {NoStop}%
\bibitem [{\citenamefont {Hu}(2024)}]{Hu:2024tnn}%
  \BibitemOpen
  \bibfield  {author} {\bibinfo {author} {\bibfnamefont {J.}~\bibnamefont {Hu}},\ }\href@noop {} {\  (\bibinfo {year} {2024})},\ \Eprint {http://arxiv.org/abs/2409.05131} {arXiv:2409.05131 [hep-ph]} \BibitemShut {NoStop}%
\bibitem [{\citenamefont {Singh}\ \emph {et~al.}(2025)\citenamefont {Singh}, \citenamefont {Bhadury}, \citenamefont {Kurian},\ and\ \citenamefont {Chandra}}]{Singh:2025wov}%
  \BibitemOpen
  \bibfield  {author} {\bibinfo {author} {\bibfnamefont {S.~K.}\ \bibnamefont {Singh}}, \bibinfo {author} {\bibfnamefont {S.}~\bibnamefont {Bhadury}}, \bibinfo {author} {\bibfnamefont {M.}~\bibnamefont {Kurian}}, \ and\ \bibinfo {author} {\bibfnamefont {V.}~\bibnamefont {Chandra}},\ }\href@noop {} {\  (\bibinfo {year} {2025})},\ \Eprint {http://arxiv.org/abs/2501.17442} {arXiv:2501.17442 [hep-ph]} \BibitemShut {NoStop}%
\bibitem [{\citenamefont {Mukherjee}\ \emph {et~al.}(2025)\citenamefont {Mukherjee}, \citenamefont {Bhadury},\ and\ \citenamefont {Singha}}]{Mukherjee:2025dqp}%
  \BibitemOpen
  \bibfield  {author} {\bibinfo {author} {\bibfnamefont {A.}~\bibnamefont {Mukherjee}}, \bibinfo {author} {\bibfnamefont {S.}~\bibnamefont {Bhadury}}, \ and\ \bibinfo {author} {\bibfnamefont {P.}~\bibnamefont {Singha}},\ }\href@noop {} {\  (\bibinfo {year} {2025})},\ \Eprint {http://arxiv.org/abs/2504.11572} {arXiv:2504.11572 [hep-ph]} \BibitemShut {NoStop}%
\bibitem [{\citenamefont {Blaizot}\ and\ \citenamefont {Yan}(2021)}]{Blaizot:2021cdv}%
  \BibitemOpen
  \bibfield  {author} {\bibinfo {author} {\bibfnamefont {J.-P.}\ \bibnamefont {Blaizot}}\ and\ \bibinfo {author} {\bibfnamefont {L.}~\bibnamefont {Yan}},\ }\href {\doibase 10.1103/PhysRevC.104.055201} {\bibfield  {journal} {\bibinfo  {journal} {Phys. Rev. C}\ }\textbf {\bibinfo {volume} {104}},\ \bibinfo {pages} {055201} (\bibinfo {year} {2021})},\ \Eprint {http://arxiv.org/abs/2106.10508} {arXiv:2106.10508 [nucl-th]} \BibitemShut {NoStop}%
\bibitem [{\citenamefont {Aniceto}\ \emph {et~al.}(2025)\citenamefont {Aniceto}, \citenamefont {Noronha},\ and\ \citenamefont {Spal\'\i{}nski}}]{Aniceto:2024pyc}%
  \BibitemOpen
  \bibfield  {author} {\bibinfo {author} {\bibfnamefont {I.}~\bibnamefont {Aniceto}}, \bibinfo {author} {\bibfnamefont {J.}~\bibnamefont {Noronha}}, \ and\ \bibinfo {author} {\bibfnamefont {M.~l.}\ \bibnamefont {Spal\'\i{}nski}},\ }\href {\doibase 10.1103/PhysRevD.111.076025} {\bibfield  {journal} {\bibinfo  {journal} {Phys. Rev. D}\ }\textbf {\bibinfo {volume} {111}},\ \bibinfo {pages} {076025} (\bibinfo {year} {2025})},\ \Eprint {http://arxiv.org/abs/2401.06750} {arXiv:2401.06750 [nucl-th]} \BibitemShut {NoStop}%
\bibitem [{\citenamefont {Ambrus}\ \emph {et~al.}(2022)\citenamefont {Ambrus}, \citenamefont {Moln\'ar},\ and\ \citenamefont {Rischke}}]{Ambrus:2022vif}%
  \BibitemOpen
  \bibfield  {author} {\bibinfo {author} {\bibfnamefont {V.~E.}\ \bibnamefont {Ambrus}}, \bibinfo {author} {\bibfnamefont {E.}~\bibnamefont {Moln\'ar}}, \ and\ \bibinfo {author} {\bibfnamefont {D.~H.}\ \bibnamefont {Rischke}},\ }\href {\doibase 10.1103/PhysRevD.106.076005} {\bibfield  {journal} {\bibinfo  {journal} {Phys. Rev. D}\ }\textbf {\bibinfo {volume} {106}},\ \bibinfo {pages} {076005} (\bibinfo {year} {2022})},\ \Eprint {http://arxiv.org/abs/2207.05670} {arXiv:2207.05670 [nucl-th]} \BibitemShut {NoStop}%
\bibitem [{\citenamefont {Jaiswal}\ \emph {et~al.}(2022)\citenamefont {Jaiswal}, \citenamefont {Blaizot}, \citenamefont {Bhalerao}, \citenamefont {Chen}, \citenamefont {Jaiswal},\ and\ \citenamefont {Yan}}]{Jaiswal:2022udf}%
  \BibitemOpen
  \bibfield  {author} {\bibinfo {author} {\bibfnamefont {S.}~\bibnamefont {Jaiswal}}, \bibinfo {author} {\bibfnamefont {J.-P.}\ \bibnamefont {Blaizot}}, \bibinfo {author} {\bibfnamefont {R.~S.}\ \bibnamefont {Bhalerao}}, \bibinfo {author} {\bibfnamefont {Z.}~\bibnamefont {Chen}}, \bibinfo {author} {\bibfnamefont {A.}~\bibnamefont {Jaiswal}}, \ and\ \bibinfo {author} {\bibfnamefont {L.}~\bibnamefont {Yan}},\ }\href {\doibase 10.1103/PhysRevC.106.044912} {\bibfield  {journal} {\bibinfo  {journal} {Phys. Rev. C}\ }\textbf {\bibinfo {volume} {106}},\ \bibinfo {pages} {044912} (\bibinfo {year} {2022})},\ \Eprint {http://arxiv.org/abs/2208.02750} {arXiv:2208.02750 [nucl-th]} \BibitemShut {NoStop}%
\bibitem [{\citenamefont {Ambru\c{s}}\ \emph {et~al.}(2024)\citenamefont {Ambru\c{s}}, \citenamefont {Moln\'ar},\ and\ \citenamefont {Rischke}}]{Ambrus:2023qcl}%
  \BibitemOpen
  \bibfield  {author} {\bibinfo {author} {\bibfnamefont {V.~E.}\ \bibnamefont {Ambru\c{s}}}, \bibinfo {author} {\bibfnamefont {E.}~\bibnamefont {Moln\'ar}}, \ and\ \bibinfo {author} {\bibfnamefont {D.~H.}\ \bibnamefont {Rischke}},\ }\href {\doibase 10.1103/PhysRevD.109.076001} {\bibfield  {journal} {\bibinfo  {journal} {Phys. Rev. D}\ }\textbf {\bibinfo {volume} {109}},\ \bibinfo {pages} {076001} (\bibinfo {year} {2024})},\ \Eprint {http://arxiv.org/abs/2311.00351} {arXiv:2311.00351 [nucl-th]} \BibitemShut {NoStop}%
\bibitem [{\citenamefont {Ochsenfeld}\ and\ \citenamefont {Schlichting}(2023)}]{Ochsenfeld:2023wxz}%
  \BibitemOpen
  \bibfield  {author} {\bibinfo {author} {\bibfnamefont {S.}~\bibnamefont {Ochsenfeld}}\ and\ \bibinfo {author} {\bibfnamefont {S.}~\bibnamefont {Schlichting}},\ }\href {\doibase 10.1007/JHEP09(2023)186} {\bibfield  {journal} {\bibinfo  {journal} {JHEP}\ }\textbf {\bibinfo {volume} {09}},\ \bibinfo {pages} {186} (\bibinfo {year} {2023})},\ \Eprint {http://arxiv.org/abs/2308.04491} {arXiv:2308.04491 [hep-th]} \BibitemShut {NoStop}%
\bibitem [{\citenamefont {Bajec}\ \emph {et~al.}(2024)\citenamefont {Bajec}, \citenamefont {Grozdanov},\ and\ \citenamefont {Soloviev}}]{Bajec:2024jez}%
  \BibitemOpen
  \bibfield  {author} {\bibinfo {author} {\bibfnamefont {M.}~\bibnamefont {Bajec}}, \bibinfo {author} {\bibfnamefont {S.}~\bibnamefont {Grozdanov}}, \ and\ \bibinfo {author} {\bibfnamefont {A.}~\bibnamefont {Soloviev}},\ }\href {\doibase 10.1007/JHEP08(2024)065} {\bibfield  {journal} {\bibinfo  {journal} {JHEP}\ }\textbf {\bibinfo {volume} {08}},\ \bibinfo {pages} {065} (\bibinfo {year} {2024})},\ \Eprint {http://arxiv.org/abs/2403.17769} {arXiv:2403.17769 [hep-th]} \BibitemShut {NoStop}%
\bibitem [{\citenamefont {Abbasi}\ and\ \citenamefont {Rischke}(2025)}]{Abbasi:2024pwz}%
  \BibitemOpen
  \bibfield  {author} {\bibinfo {author} {\bibfnamefont {N.}~\bibnamefont {Abbasi}}\ and\ \bibinfo {author} {\bibfnamefont {D.~H.}\ \bibnamefont {Rischke}},\ }\href {\doibase 10.1007/JHEP05(2025)241} {\bibfield  {journal} {\bibinfo  {journal} {JHEP}\ }\textbf {\bibinfo {volume} {05}},\ \bibinfo {pages} {241} (\bibinfo {year} {2025})},\ \Eprint {http://arxiv.org/abs/2410.07929} {arXiv:2410.07929 [hep-th]} \BibitemShut {NoStop}%
\bibitem [{\citenamefont {Rocha}\ and\ \citenamefont {Denicol}(2025)}]{Rocha:2025rkl}%
  \BibitemOpen
  \bibfield  {author} {\bibinfo {author} {\bibfnamefont {G.~S.}\ \bibnamefont {Rocha}}\ and\ \bibinfo {author} {\bibfnamefont {G.~S.}\ \bibnamefont {Denicol}},\ }\href@noop {} {\  (\bibinfo {year} {2025})},\ \Eprint {http://arxiv.org/abs/2505.14823} {arXiv:2505.14823 [nucl-th]} \BibitemShut {NoStop}%
\bibitem [{\citenamefont {Dusling}\ and\ \citenamefont {Sch\"afer}(2012)}]{Dusling:2011fd}%
  \BibitemOpen
  \bibfield  {author} {\bibinfo {author} {\bibfnamefont {K.}~\bibnamefont {Dusling}}\ and\ \bibinfo {author} {\bibfnamefont {T.}~\bibnamefont {Sch\"afer}},\ }\href {\doibase 10.1103/PhysRevC.85.044909} {\bibfield  {journal} {\bibinfo  {journal} {Phys. Rev. C}\ }\textbf {\bibinfo {volume} {85}},\ \bibinfo {pages} {044909} (\bibinfo {year} {2012})},\ \Eprint {http://arxiv.org/abs/1109.5181} {arXiv:1109.5181 [hep-ph]} \BibitemShut {NoStop}%
\bibitem [{\citenamefont {Das}\ \emph {et~al.}(2020)\citenamefont {Das}, \citenamefont {Mishra},\ and\ \citenamefont {Mohapatra}}]{Das:2020beh}%
  \BibitemOpen
  \bibfield  {author} {\bibinfo {author} {\bibfnamefont {A.}~\bibnamefont {Das}}, \bibinfo {author} {\bibfnamefont {H.}~\bibnamefont {Mishra}}, \ and\ \bibinfo {author} {\bibfnamefont {R.~K.}\ \bibnamefont {Mohapatra}},\ }\href {\doibase 10.1103/PhysRevD.102.014030} {\bibfield  {journal} {\bibinfo  {journal} {Phys. Rev. D}\ }\textbf {\bibinfo {volume} {102}},\ \bibinfo {pages} {014030} (\bibinfo {year} {2020})},\ \Eprint {http://arxiv.org/abs/2004.04665} {arXiv:2004.04665 [hep-ph]} \BibitemShut {NoStop}%
\bibitem [{\citenamefont {Zebarjadi}\ \emph {et~al.}(2021)\citenamefont {Zebarjadi}, \citenamefont {Rezaei}, \citenamefont {Akhanda},\ and\ \citenamefont {Esfarjani}}]{PhysRevB.103.144404}%
  \BibitemOpen
  \bibfield  {author} {\bibinfo {author} {\bibfnamefont {M.}~\bibnamefont {Zebarjadi}}, \bibinfo {author} {\bibfnamefont {S.~E.}\ \bibnamefont {Rezaei}}, \bibinfo {author} {\bibfnamefont {M.~S.}\ \bibnamefont {Akhanda}}, \ and\ \bibinfo {author} {\bibfnamefont {K.}~\bibnamefont {Esfarjani}},\ }\href {\doibase 10.1103/PhysRevB.103.144404} {\bibfield  {journal} {\bibinfo  {journal} {Phys. Rev. B}\ }\textbf {\bibinfo {volume} {103}},\ \bibinfo {pages} {144404} (\bibinfo {year} {2021})}\BibitemShut {NoStop}%
\bibitem [{\citenamefont {Romatschke}(2012)}]{Romatschke:2011qp}%
  \BibitemOpen
  \bibfield  {author} {\bibinfo {author} {\bibfnamefont {P.}~\bibnamefont {Romatschke}},\ }\href {\doibase 10.1103/PhysRevD.85.065012} {\bibfield  {journal} {\bibinfo  {journal} {Phys. Rev. D}\ }\textbf {\bibinfo {volume} {85}},\ \bibinfo {pages} {065012} (\bibinfo {year} {2012})},\ \Eprint {http://arxiv.org/abs/1108.5561} {arXiv:1108.5561 [gr-qc]} \BibitemShut {NoStop}%
\bibitem [{\citenamefont {Hataei}\ \emph {et~al.}(2025)\citenamefont {Hataei}, \citenamefont {Heydari},\ and\ \citenamefont {Taghinavaz}}]{Hataei:2025mqf}%
  \BibitemOpen
  \bibfield  {author} {\bibinfo {author} {\bibfnamefont {A.}~\bibnamefont {Hataei}}, \bibinfo {author} {\bibfnamefont {R.}~\bibnamefont {Heydari}}, \ and\ \bibinfo {author} {\bibfnamefont {F.}~\bibnamefont {Taghinavaz}},\ }\href@noop {} {\  (\bibinfo {year} {2025})},\ \Eprint {http://arxiv.org/abs/2504.14591} {arXiv:2504.14591 [nucl-th]} \BibitemShut {NoStop}%
\bibitem [{\citenamefont {Lin}\ \emph {et~al.}(2025)\citenamefont {Lin}, \citenamefont {Sun}, \citenamefont {Wu},\ and\ \citenamefont {Hu}}]{Lin:2025ehr}%
  \BibitemOpen
  \bibfield  {author} {\bibinfo {author} {\bibfnamefont {X.}~\bibnamefont {Lin}}, \bibinfo {author} {\bibfnamefont {Q.-Z.}\ \bibnamefont {Sun}}, \bibinfo {author} {\bibfnamefont {X.-H.}\ \bibnamefont {Wu}}, \ and\ \bibinfo {author} {\bibfnamefont {J.}~\bibnamefont {Hu}},\ }\href@noop {} {\  (\bibinfo {year} {2025})},\ \Eprint {http://arxiv.org/abs/2505.04444} {arXiv:2505.04444 [hep-ph]} \BibitemShut {NoStop}%
\bibitem [{\citenamefont {Grozdanov}\ \emph {et~al.}(2019)\citenamefont {Grozdanov}, \citenamefont {Lucas},\ and\ \citenamefont {Poovuttikul}}]{Grozdanov:2018fic}%
  \BibitemOpen
  \bibfield  {author} {\bibinfo {author} {\bibfnamefont {S.}~\bibnamefont {Grozdanov}}, \bibinfo {author} {\bibfnamefont {A.}~\bibnamefont {Lucas}}, \ and\ \bibinfo {author} {\bibfnamefont {N.}~\bibnamefont {Poovuttikul}},\ }\href {\doibase 10.1103/PhysRevD.99.086012} {\bibfield  {journal} {\bibinfo  {journal} {Phys. Rev. D}\ }\textbf {\bibinfo {volume} {99}},\ \bibinfo {pages} {086012} (\bibinfo {year} {2019})},\ \Eprint {http://arxiv.org/abs/1810.10016} {arXiv:1810.10016 [hep-th]} \BibitemShut {NoStop}%
\bibitem [{\citenamefont {Heller}\ \emph {et~al.}(2023)\citenamefont {Heller}, \citenamefont {Serantes}, \citenamefont {Spali\'nski},\ and\ \citenamefont {Withers}}]{Heller:2022ejw}%
  \BibitemOpen
  \bibfield  {author} {\bibinfo {author} {\bibfnamefont {M.~P.}\ \bibnamefont {Heller}}, \bibinfo {author} {\bibfnamefont {A.}~\bibnamefont {Serantes}}, \bibinfo {author} {\bibfnamefont {M.}~\bibnamefont {Spali\'nski}}, \ and\ \bibinfo {author} {\bibfnamefont {B.}~\bibnamefont {Withers}},\ }\href {\doibase 10.1103/PhysRevLett.130.261601} {\bibfield  {journal} {\bibinfo  {journal} {Phys. Rev. Lett.}\ }\textbf {\bibinfo {volume} {130}},\ \bibinfo {pages} {261601} (\bibinfo {year} {2023})},\ \Eprint {http://arxiv.org/abs/2212.07434} {arXiv:2212.07434 [hep-th]} \BibitemShut {NoStop}%
\bibitem [{\citenamefont {Heller}\ \emph {et~al.}(2024)\citenamefont {Heller}, \citenamefont {Serantes}, \citenamefont {Spali\'nski},\ and\ \citenamefont {Withers}}]{Heller:2023jtd}%
  \BibitemOpen
  \bibfield  {author} {\bibinfo {author} {\bibfnamefont {M.~P.}\ \bibnamefont {Heller}}, \bibinfo {author} {\bibfnamefont {A.}~\bibnamefont {Serantes}}, \bibinfo {author} {\bibfnamefont {M.}~\bibnamefont {Spali\'nski}}, \ and\ \bibinfo {author} {\bibfnamefont {B.}~\bibnamefont {Withers}},\ }\href {\doibase 10.1038/s41567-024-02635-5} {\bibfield  {journal} {\bibinfo  {journal} {Nature Phys.}\ }\textbf {\bibinfo {volume} {20}},\ \bibinfo {pages} {1948} (\bibinfo {year} {2024})},\ \Eprint {http://arxiv.org/abs/2305.07703} {arXiv:2305.07703 [hep-th]} \BibitemShut {NoStop}%
\bibitem [{\citenamefont {Gavassino}\ \emph {et~al.}(2024)\citenamefont {Gavassino}, \citenamefont {Disconzi},\ and\ \citenamefont {Noronha}}]{Gavassino:2023mad}%
  \BibitemOpen
  \bibfield  {author} {\bibinfo {author} {\bibfnamefont {L.}~\bibnamefont {Gavassino}}, \bibinfo {author} {\bibfnamefont {M.~M.}\ \bibnamefont {Disconzi}}, \ and\ \bibinfo {author} {\bibfnamefont {J.}~\bibnamefont {Noronha}},\ }\href {\doibase 10.1103/PhysRevLett.132.162301} {\bibfield  {journal} {\bibinfo  {journal} {Phys. Rev. Lett.}\ }\textbf {\bibinfo {volume} {132}},\ \bibinfo {pages} {162301} (\bibinfo {year} {2024})},\ \Eprint {http://arxiv.org/abs/2307.05987} {arXiv:2307.05987 [hep-th]} \BibitemShut {NoStop}%
\bibitem [{\citenamefont {Hartnoll}\ and\ \citenamefont {Kumar}(2005)}]{Hartnoll:2005ju}%
  \BibitemOpen
  \bibfield  {author} {\bibinfo {author} {\bibfnamefont {S.~A.}\ \bibnamefont {Hartnoll}}\ and\ \bibinfo {author} {\bibfnamefont {S.~P.}\ \bibnamefont {Kumar}},\ }\href {\doibase 10.1088/1126-6708/2005/12/036} {\bibfield  {journal} {\bibinfo  {journal} {JHEP}\ }\textbf {\bibinfo {volume} {12}},\ \bibinfo {pages} {036} (\bibinfo {year} {2005})},\ \Eprint {http://arxiv.org/abs/hep-th/0508092} {arXiv:hep-th/0508092} \BibitemShut {NoStop}%
\bibitem [{\citenamefont {Rocha}\ \emph {et~al.}(2024)\citenamefont {Rocha}, \citenamefont {Danhoni}, \citenamefont {Ingles}, \citenamefont {Denicol},\ and\ \citenamefont {Noronha}}]{Rocha:2024cge}%
  \BibitemOpen
  \bibfield  {author} {\bibinfo {author} {\bibfnamefont {G.~S.}\ \bibnamefont {Rocha}}, \bibinfo {author} {\bibfnamefont {I.}~\bibnamefont {Danhoni}}, \bibinfo {author} {\bibfnamefont {K.}~\bibnamefont {Ingles}}, \bibinfo {author} {\bibfnamefont {G.~S.}\ \bibnamefont {Denicol}}, \ and\ \bibinfo {author} {\bibfnamefont {J.}~\bibnamefont {Noronha}},\ }\href {\doibase 10.1103/PhysRevD.110.076003} {\bibfield  {journal} {\bibinfo  {journal} {Phys. Rev. D}\ }\textbf {\bibinfo {volume} {110}},\ \bibinfo {pages} {076003} (\bibinfo {year} {2024})},\ \Eprint {http://arxiv.org/abs/2404.04679} {arXiv:2404.04679 [nucl-th]} \BibitemShut {NoStop}%
\bibitem [{\citenamefont {Casalderrey-Solana}\ \emph {et~al.}(2018)\citenamefont {Casalderrey-Solana}, \citenamefont {Grozdanov},\ and\ \citenamefont {Starinets}}]{Casalderrey-Solana:2018rle}%
  \BibitemOpen
  \bibfield  {author} {\bibinfo {author} {\bibfnamefont {J.}~\bibnamefont {Casalderrey-Solana}}, \bibinfo {author} {\bibfnamefont {S.}~\bibnamefont {Grozdanov}}, \ and\ \bibinfo {author} {\bibfnamefont {A.~O.}\ \bibnamefont {Starinets}},\ }\href {\doibase 10.1103/PhysRevLett.121.191603} {\bibfield  {journal} {\bibinfo  {journal} {Phys. Rev. Lett.}\ }\textbf {\bibinfo {volume} {121}},\ \bibinfo {pages} {191603} (\bibinfo {year} {2018})},\ \Eprint {http://arxiv.org/abs/1806.10997} {arXiv:1806.10997 [hep-th]} \BibitemShut {NoStop}%
\bibitem [{\citenamefont {Grozdanov}\ and\ \citenamefont {Starinets}(2019)}]{Grozdanov:2018gfx}%
  \BibitemOpen
  \bibfield  {author} {\bibinfo {author} {\bibfnamefont {S.}~\bibnamefont {Grozdanov}}\ and\ \bibinfo {author} {\bibfnamefont {A.~O.}\ \bibnamefont {Starinets}},\ }\href {\doibase 10.1007/JHEP04(2019)080} {\bibfield  {journal} {\bibinfo  {journal} {JHEP}\ }\textbf {\bibinfo {volume} {04}},\ \bibinfo {pages} {080} (\bibinfo {year} {2019})},\ \Eprint {http://arxiv.org/abs/1812.09288} {arXiv:1812.09288 [hep-th]} \BibitemShut {NoStop}%
\bibitem [{\citenamefont {Gursoy}\ and\ \citenamefont {Kiritsis}(2008)}]{Gursoy:2007cb}%
  \BibitemOpen
  \bibfield  {author} {\bibinfo {author} {\bibfnamefont {U.}~\bibnamefont {Gursoy}}\ and\ \bibinfo {author} {\bibfnamefont {E.}~\bibnamefont {Kiritsis}},\ }\href {\doibase 10.1088/1126-6708/2008/02/032} {\bibfield  {journal} {\bibinfo  {journal} {JHEP}\ }\textbf {\bibinfo {volume} {02}},\ \bibinfo {pages} {032} (\bibinfo {year} {2008})},\ \Eprint {http://arxiv.org/abs/0707.1324} {arXiv:0707.1324 [hep-th]} \BibitemShut {NoStop}%
\bibitem [{\citenamefont {Gursoy}\ \emph {et~al.}(2008)\citenamefont {Gursoy}, \citenamefont {Kiritsis},\ and\ \citenamefont {Nitti}}]{Gursoy:2007er}%
  \BibitemOpen
  \bibfield  {author} {\bibinfo {author} {\bibfnamefont {U.}~\bibnamefont {Gursoy}}, \bibinfo {author} {\bibfnamefont {E.}~\bibnamefont {Kiritsis}}, \ and\ \bibinfo {author} {\bibfnamefont {F.}~\bibnamefont {Nitti}},\ }\href {\doibase 10.1088/1126-6708/2008/02/019} {\bibfield  {journal} {\bibinfo  {journal} {JHEP}\ }\textbf {\bibinfo {volume} {02}},\ \bibinfo {pages} {019} (\bibinfo {year} {2008})},\ \Eprint {http://arxiv.org/abs/0707.1349} {arXiv:0707.1349 [hep-th]} \BibitemShut {NoStop}%
\bibitem [{\citenamefont {Betzios}\ \emph {et~al.}(2018)\citenamefont {Betzios}, \citenamefont {G\"ursoy}, \citenamefont {J\"arvinen},\ and\ \citenamefont {Policastro}}]{Betzios:2017dol}%
  \BibitemOpen
  \bibfield  {author} {\bibinfo {author} {\bibfnamefont {P.}~\bibnamefont {Betzios}}, \bibinfo {author} {\bibfnamefont {U.}~\bibnamefont {G\"ursoy}}, \bibinfo {author} {\bibfnamefont {M.}~\bibnamefont {J\"arvinen}}, \ and\ \bibinfo {author} {\bibfnamefont {G.}~\bibnamefont {Policastro}},\ }\href {\doibase 10.1103/PhysRevD.97.081901} {\bibfield  {journal} {\bibinfo  {journal} {Phys. Rev. D}\ }\textbf {\bibinfo {volume} {97}},\ \bibinfo {pages} {081901} (\bibinfo {year} {2018})},\ \Eprint {http://arxiv.org/abs/1708.02252} {arXiv:1708.02252 [hep-th]} \BibitemShut {NoStop}%
\bibitem [{\citenamefont {Gradshteyn}\ \emph {et~al.}(2015)\citenamefont {Gradshteyn}, \citenamefont {Ryzhik}, \citenamefont {Zwillinger},\ and\ \citenamefont {Moll}}]{Gradshteyn:1702455}%
  \BibitemOpen
  \bibfield  {author} {\bibinfo {author} {\bibfnamefont {I.~S.}\ \bibnamefont {Gradshteyn}}, \bibinfo {author} {\bibfnamefont {I.~M.}\ \bibnamefont {Ryzhik}}, \bibinfo {author} {\bibfnamefont {D.}~\bibnamefont {Zwillinger}}, \ and\ \bibinfo {author} {\bibfnamefont {V.}~\bibnamefont {Moll}},\ }\href {\doibase 0123849330} {\emph {\bibinfo {title} {{Table of integrals, series, and products; 8th ed.}}}}\ (\bibinfo  {publisher} {Academic Press},\ \bibinfo {address} {Amsterdam},\ \bibinfo {year} {2015})\BibitemShut {NoStop}%
\bibitem [{\citenamefont {Kovtun}(2012)}]{Kovtun:2012rj}%
  \BibitemOpen
  \bibfield  {author} {\bibinfo {author} {\bibfnamefont {P.}~\bibnamefont {Kovtun}},\ }\href {\doibase 10.1088/1751-8113/45/47/473001} {\bibfield  {journal} {\bibinfo  {journal} {J. Phys. A}\ }\textbf {\bibinfo {volume} {45}},\ \bibinfo {pages} {473001} (\bibinfo {year} {2012})},\ \Eprint {http://arxiv.org/abs/1205.5040} {arXiv:1205.5040 [hep-th]} \BibitemShut {NoStop}%
\bibitem [{\citenamefont {Hattori}\ \emph {et~al.}(2022)\citenamefont {Hattori}, \citenamefont {Hongo},\ and\ \citenamefont {Huang}}]{Hattori:2022hyo}%
  \BibitemOpen
  \bibfield  {author} {\bibinfo {author} {\bibfnamefont {K.}~\bibnamefont {Hattori}}, \bibinfo {author} {\bibfnamefont {M.}~\bibnamefont {Hongo}}, \ and\ \bibinfo {author} {\bibfnamefont {X.-G.}\ \bibnamefont {Huang}},\ }\href {\doibase 10.3390/sym14091851} {\bibfield  {journal} {\bibinfo  {journal} {Symmetry}\ }\textbf {\bibinfo {volume} {14}},\ \bibinfo {pages} {1851} (\bibinfo {year} {2022})},\ \Eprint {http://arxiv.org/abs/2207.12794} {arXiv:2207.12794 [hep-th]} \BibitemShut {NoStop}%
\bibitem [{\citenamefont {Cercignani}\ and\ \citenamefont {Kremer}(2002)}]{ckbook}%
  \BibitemOpen
  \bibfield  {author} {\bibinfo {author} {\bibfnamefont {C.}~\bibnamefont {Cercignani}}\ and\ \bibinfo {author} {\bibfnamefont {G.~M.}\ \bibnamefont {Kremer}},\ }\href@noop {} {\emph {\bibinfo {title} {{The Relativistic Boltzmann Equation: Theory and Applications}}}}\ (\bibinfo  {publisher} {{Birkhäuser Basel}},\ \bibinfo {year} {2002})\BibitemShut {NoStop}%
\bibitem [{\citenamefont {Hartnoll}\ \emph {et~al.}(2007)\citenamefont {Hartnoll}, \citenamefont {Kovtun}, \citenamefont {Muller},\ and\ \citenamefont {Sachdev}}]{Hartnoll:2007ih}%
  \BibitemOpen
  \bibfield  {author} {\bibinfo {author} {\bibfnamefont {S.~A.}\ \bibnamefont {Hartnoll}}, \bibinfo {author} {\bibfnamefont {P.~K.}\ \bibnamefont {Kovtun}}, \bibinfo {author} {\bibfnamefont {M.}~\bibnamefont {Muller}}, \ and\ \bibinfo {author} {\bibfnamefont {S.}~\bibnamefont {Sachdev}},\ }\href {\doibase 10.1103/PhysRevB.76.144502} {\bibfield  {journal} {\bibinfo  {journal} {Phys. Rev. B}\ }\textbf {\bibinfo {volume} {76}},\ \bibinfo {pages} {144502} (\bibinfo {year} {2007})},\ \Eprint {http://arxiv.org/abs/0706.3215} {arXiv:0706.3215 [cond-mat.str-el]} \BibitemShut {NoStop}%
\bibitem [{\citenamefont {Hartnoll}\ \emph {et~al.}(2016)\citenamefont {Hartnoll}, \citenamefont {Lucas},\ and\ \citenamefont {Sachdev}}]{Hartnoll:2016apf}%
  \BibitemOpen
  \bibfield  {author} {\bibinfo {author} {\bibfnamefont {S.~A.}\ \bibnamefont {Hartnoll}}, \bibinfo {author} {\bibfnamefont {A.}~\bibnamefont {Lucas}}, \ and\ \bibinfo {author} {\bibfnamefont {S.}~\bibnamefont {Sachdev}},\ }\href@noop {} {\  (\bibinfo {year} {2016})},\ \Eprint {http://arxiv.org/abs/1612.07324} {arXiv:1612.07324 [hep-th]} \BibitemShut {NoStop}%
\bibitem [{\citenamefont {Kurkela}\ \emph {et~al.}(2019{\natexlab{b}})\citenamefont {Kurkela}, \citenamefont {Wiedemann},\ and\ \citenamefont {Wu}}]{Kurkela:2018qeb}%
  \BibitemOpen
  \bibfield  {author} {\bibinfo {author} {\bibfnamefont {A.}~\bibnamefont {Kurkela}}, \bibinfo {author} {\bibfnamefont {U.~A.}\ \bibnamefont {Wiedemann}}, \ and\ \bibinfo {author} {\bibfnamefont {B.}~\bibnamefont {Wu}},\ }\href {\doibase 10.1140/epjc/s10052-019-7262-x} {\bibfield  {journal} {\bibinfo  {journal} {Eur. Phys. J. C}\ }\textbf {\bibinfo {volume} {79}},\ \bibinfo {pages} {759} (\bibinfo {year} {2019}{\natexlab{b}})},\ \Eprint {http://arxiv.org/abs/1805.04081} {arXiv:1805.04081 [hep-ph]} \BibitemShut {NoStop}%
\bibitem [{\citenamefont {Grozdanov}\ and\ \citenamefont {Soloviev}(2024)}]{Grozdanov:2024fxr}%
  \BibitemOpen
  \bibfield  {author} {\bibinfo {author} {\bibfnamefont {S.}~\bibnamefont {Grozdanov}}\ and\ \bibinfo {author} {\bibfnamefont {A.}~\bibnamefont {Soloviev}},\ }\href@noop {} {\  (\bibinfo {year} {2024})},\ \Eprint {http://arxiv.org/abs/2501.00099} {arXiv:2501.00099 [hep-th]} \BibitemShut {NoStop}%
\bibitem [{\citenamefont {Bazow}\ \emph {et~al.}(2016)\citenamefont {Bazow}, \citenamefont {Denicol}, \citenamefont {Heinz}, \citenamefont {Martinez},\ and\ \citenamefont {Noronha}}]{PhysRevLett.116.022301}%
  \BibitemOpen
  \bibfield  {author} {\bibinfo {author} {\bibfnamefont {D.}~\bibnamefont {Bazow}}, \bibinfo {author} {\bibfnamefont {G.~S.}\ \bibnamefont {Denicol}}, \bibinfo {author} {\bibfnamefont {U.}~\bibnamefont {Heinz}}, \bibinfo {author} {\bibfnamefont {M.}~\bibnamefont {Martinez}}, \ and\ \bibinfo {author} {\bibfnamefont {J.}~\bibnamefont {Noronha}},\ }\href {\doibase 10.1103/PhysRevLett.116.022301} {\bibfield  {journal} {\bibinfo  {journal} {Phys. Rev. Lett.}\ }\textbf {\bibinfo {volume} {116}},\ \bibinfo {pages} {022301} (\bibinfo {year} {2016})}\BibitemShut {NoStop}%
\bibitem [{\citenamefont {Kurkela}\ \emph {et~al.}(2018)\citenamefont {Kurkela}, \citenamefont {Mukhopadhyay}, \citenamefont {Preis}, \citenamefont {Rebhan},\ and\ \citenamefont {Soloviev}}]{Kurkela:2018dku}%
  \BibitemOpen
  \bibfield  {author} {\bibinfo {author} {\bibfnamefont {A.}~\bibnamefont {Kurkela}}, \bibinfo {author} {\bibfnamefont {A.}~\bibnamefont {Mukhopadhyay}}, \bibinfo {author} {\bibfnamefont {F.}~\bibnamefont {Preis}}, \bibinfo {author} {\bibfnamefont {A.}~\bibnamefont {Rebhan}}, \ and\ \bibinfo {author} {\bibfnamefont {A.}~\bibnamefont {Soloviev}},\ }\href {\doibase 10.1007/JHEP08(2018)054} {\bibfield  {journal} {\bibinfo  {journal} {JHEP}\ }\textbf {\bibinfo {volume} {08}},\ \bibinfo {pages} {054} (\bibinfo {year} {2018})},\ \Eprint {http://arxiv.org/abs/1805.05213} {arXiv:1805.05213 [hep-ph]} \BibitemShut {NoStop}%
\bibitem [{\citenamefont {Abramowitz}\ and\ \citenamefont {Stegun}(1964)}]{Abramowitz1964}%
  \BibitemOpen
  \bibfield  {author} {\bibinfo {author} {\bibfnamefont {M.}~\bibnamefont {Abramowitz}}\ and\ \bibinfo {author} {\bibfnamefont {I.}~\bibnamefont {Stegun}},\ }\href@noop {} {\emph {\bibinfo {title} {Handbook of mathematical functions with formulas, graphs, and mathematical tables}}}\ (\bibinfo  {publisher} {Dover},\ \bibinfo {address} {New York},\ \bibinfo {year} {1964})\BibitemShut {NoStop}%
\bibitem [{\citenamefont {Ahlfors}()}]{Ahlfors1966}%
  \BibitemOpen
  \bibfield  {author} {\bibinfo {author} {\bibfnamefont {L.~V.}\ \bibnamefont {Ahlfors}},\ }\href@noop {} {\emph {\bibinfo {title} {Complex Analysis}}},\ \bibinfo {edition} {2nd}\ ed.\ (\bibinfo  {publisher} {McGraw-Hill Book Company})\BibitemShut {NoStop}%
\end{thebibliography}%
\bibliographystyle{apsrev4-1}

\end{document}